\documentclass{aa} 
\usepackage[varg]{txfonts}
\usepackage{epsfig}
\usepackage{natbib}
\bibpunct{(}{)}{;}{a}{}{,} 

\usepackage{graphicx}
\usepackage[]{hyperref}

\usepackage{siunitx}
\usepackage{bm}

\begin{document}

\title{A global view of shocked plasma in the supernova remnant \\ Puppis A provided by SRG/eROSITA}

\author{Martin G.~F.~Mayer\thanks{\email{mmayer@mpe.mpg.de}}\inst{1} \and Werner Becker\inst{1,2} \and Peter Predehl\inst{1} \and Manami Sasaki\inst{3} \and Michael Freyberg\inst{1}}
\institute{Max-Planck Institut f\"ur extraterrestrische Physik, Giessenbachstrasse, 85748 Garching, Germany \and 
Max-Planck Institut f\"ur Radioastronomie, Auf dem H\"ugel 69, 53121 Bonn, Germany \and Dr. Karl Remeis Observatory, Erlangen Centre for Astroparticle Physics, Friedrich-Alexander-Universit\"at Erlangen-N\"urnberg,
Sternwartstrasse 7, 96049 Bamberg, Germany}

\date{Received XXX /
Accepted YYY }

\abstract
{Puppis A is a medium-age supernova remnant (SNR), which is visible as a very bright extended X-ray source. While numerous studies have investigated individual features of the SNR, at this time, no comprehensive study of the entirety of its X-ray emission exists.} 
{Using field-scan data acquired by the SRG/eROSITA telescope during its calibration and performance verification phase, we aim to investigate the physical conditions of shocked plasma and the distribution of elements throughout Puppis A. In doing so, we take advantage of the uniform target coverage, excellent statistics, and decent spatial and spectral resolution of our data set.} 
{Using broad- and narrow-band imaging, we investigate the large-scale distribution of absorption and plasma temperature as well as that of typical emission lines. 
This approach is complemented by spatially resolved spectral analysis of the shocked plasma in Puppis A, for which we divide the SNR into around 700 distinct regions, resulting in maps of key physical quantities over its extent.} 
{We find a strong peak of foreground absorption in the southwest quadrant, which in conjunction with high temperatures at the northeast rim creates the well-known strip of hard emission crossing Puppis A.
Furthermore, using the observed distribution of ionization ages, we attempt to reconstruct the age of the shock in the individual regions. We find a quite recent shock interaction for the prominent northeast filament and ejecta knot, as well as for the outer edge of the ``bright eastern knot''. 
Finally, elemental abundance maps reveal only a single clear enhancement of the plasma with ejecta material, consistent with a previously identified region, and no obvious ejecta enrichment in the remainder of the SNR. Within this region, we confirm the spatial separation of silicon-rich ejecta from those dominated by lighter elements. 
The apparent elemental composition of this ejecta-rich region would imply an unrealistically large silicon-to-oxygen ratio when compared to the integrated yield of a core-collapse supernova. In reality, both the observed ejecta composition and their apparent distribution may be biased by the unknown location and strength of the reverse shock. 
}  
{}

\keywords{X-rays: Puppis A -- ISM: supernova remnants -- ISM: abundances -- Stars: neutron} 

\titlerunning{SRG/eROSITA spectro-imaging analysis of Puppis A}
\maketitle

\section{Introduction}
Puppis A (G260.4$-$3.4) is a nearby Galactic core-collapse supernova remnant (SNR), which is particularly noteworthy for being one of the most prominent extended X-ray sources in the sky.
It is most luminous at X-ray and infrared energies, where it appears as a deformed shell of around $56\arcmin$ diameter, with a rich substructure formed by multiple filaments, clumps, and arcs \citep{Arendt10,Dubner13}. At those wavelengths, it exhibits a strong brightness gradient from northeast to southwest \citep{Petre82}, likely caused by a density gradient in the surrounding interstellar medium (ISM).
The distance to Puppis A has most recently been estimated to $1.3\pm0.3\,\si{kpc}$ via an \ion{H}{i} absorption study \citep{Reynoso17}, which places it behind the very extended nearby Vela SNR. 
Puppis A is detected across almost the entire electromagnetic spectrum, reaching from radio up until $\si{GeV}$ energies \citep{Xin17}. However, it has so far eluded detection at $\si{TeV}$ energies, somewhat at odds with expectations \citep{HESS15}. 

Puppis A hosts the central compact object (CCO) RX J0822$-$4300. CCOs are a peculiar class of young neutron stars that have exclusively been observed in X-rays, and are notable for the fact that their emission appears to be of purely thermal nature \citep{DeLuca08}, and in some cases for their very small magnetic fields \citep{Gotthelf13}. 
Studies of this particular CCO's large proper motion, as well as  optical expansion studies of oxygen-rich ejecta knots, have provided kinematic age estimates for Puppis A in the range of $3700-4600\,\si{yr}$ \citep{Mayer20,Becker12,Winkler88}. This makes Puppis A much older than X-ray bright SNRs such as Cas A, Kepler or Tycho, but significantly younger than the neighboring Vela SNR.

\begin{figure*}
\centering
\includegraphics[width=18.0cm]{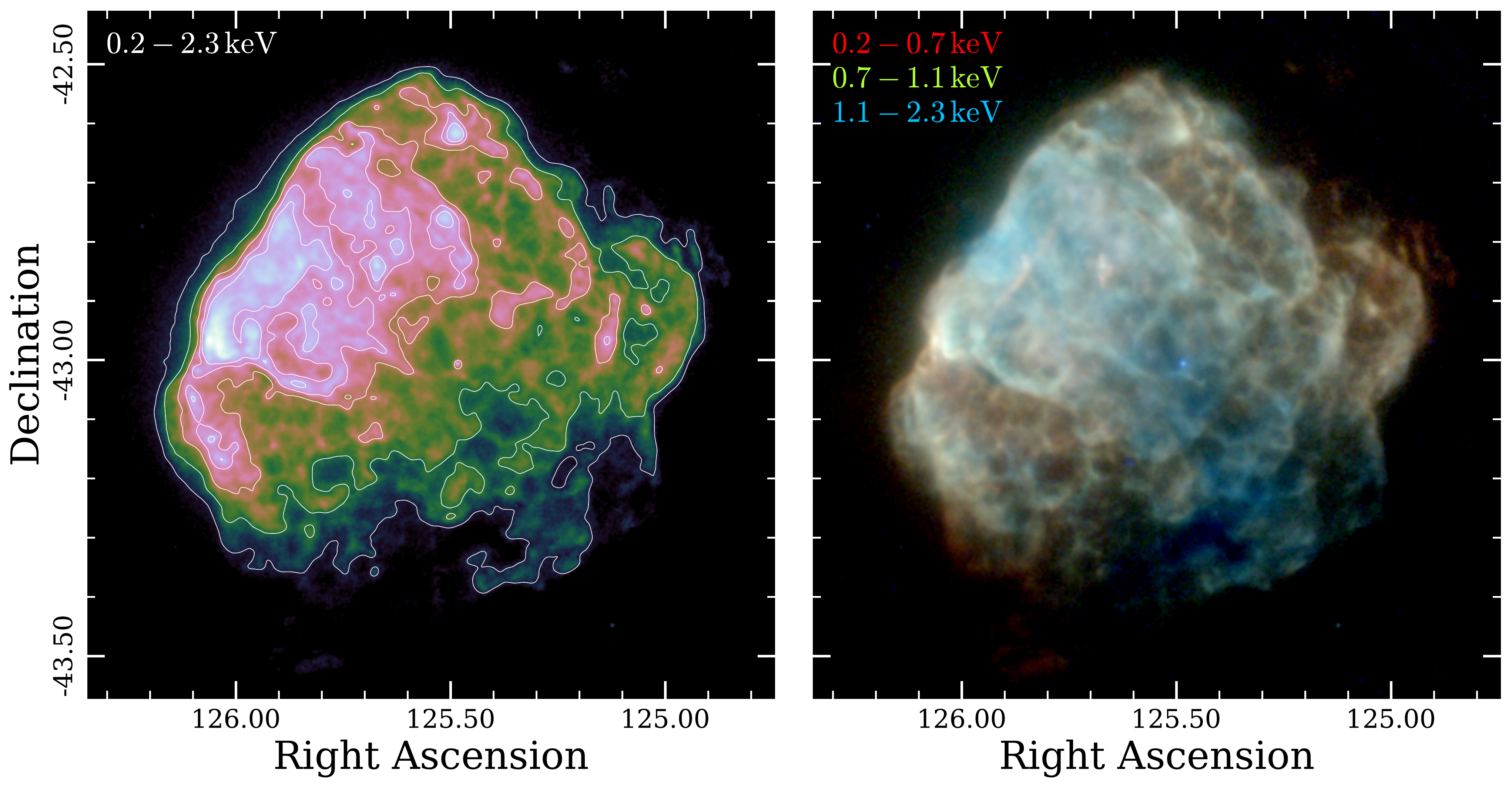} 
\caption{Single-band (left) and false-color (right) exposure-corrected images of Puppis A in the $0.2-2.3\,\si{keV}$ range as seen by eROSITA. Both images were smoothed using a Gaussian kernel of width $6\arcsec$. The color scale in this, as in all images shown in this work, is logarithmic. The contours in the left panel trace the $0.2-2.3\,\si{keV}$ emission at levels of  $1\times10^{-4}, 3\times10^{-4}, 1\times10^{-3}, 2\times10^{-3}, 3\times10^{-3}\,\si{ct.s^{-1}.arcsec^{-2}}$. For comparison, the local background level in the broad band, measured in a region around $30\arcmin$ northeast of the rim of Puppis A using the same observation, is around $1.2\times10^{-5} \,\si{ct.s^{-1}.arcsec^{-2}}$.} 
\label{PuppisImage}
\end{figure*}

Numerous previous works have identified and discussed  prominent morphological features visible in Puppis A. These include the very luminous ``bright eastern knot'' (BEK), where the supernova shock wave interacts with a concentrated ISM density enhancement, leading to an indentation of the X-ray shell \citep{Hwang05}.
Moreover, \citet{Hwang08} and \citet{Katsuda08} independently discovered the presence of a localized enhancement of metal abundances (O, Ne, Mg, Si, Fe), consistent with the presence of compact ejecta knot and a more extended ejecta-rich region somewhat north of the SNR center, whereas they found little ejecta enrichment in the remaining SNR fraction under investigation.   
Similarly, \citet{Katsuda10} found an apparent enrichment in ejecta abundances in an underionized filament running parallel to the northeast rim of Puppis A. 
Detailed insights into the radiation processes acting within the BEK became available with {\it XMM-Newton} reflection grating spectroscopy. Via analysis of line ratios, tensions with a purely thermal plasma emission model were revealed, which can possibly be relieved when including charge exchange processes \citep{Katsuda12}. 
Using the same instrument, \citet{Katsuda13} investigated the dynamics of the ejecta knot mentioned above, finding a radial velocity around $1500\,\si{km.s^{-1}}$, and providing a rare constraint on the oxygen temperature at $\lesssim 30\,\si{keV}$.

The most sensitive and complete X-ray-view of the entirety of Puppis A has been compiled by \citet{Dubner13}, who combined numerous pointed observations of {\it XMM-Newton} and {\it Chandra} to obtain a remarkable level of detail in a broad-band mosaic image of Puppis A. \citet{Luna16} used this data set to introduce an interesting new method for creating spectral extraction regions. They briefly discussed their findings on temperature structure, foreground absorption, and elemental abundances, however only using a relatively coarse spatial resolution in the displayed maps. At this time, no comprehensive, spatially resolved spectral analysis of the entire Puppis A SNR has been carried out using data from only a single instrument. 

eROSITA \citep{Predehl21} is the soft X-ray instrument aboard the German-Russian Spectrum-Roentgen-Gamma mission \citep{Sunyaev21}. It consists of seven identical X-ray imaging telescope modules (TMs) with a field of view with a diameter around one degree, which are sensitive to X-ray emission in the $0.2-10.0 \,\si{keV}$ band. Its main task is the performance of eight consecutive all-sky surveys, which cumulatively will be around 25 times deeper in the $0.2-2.3\,\si{keV}$ band than the ROSAT all-sky survey, and achieve an average spatial resolution around $26\arcsec$ \citep{Merloni12,Predehl21}. 

In this work, we use an early eROSITA calibration observation to conduct detailed spectro-imaging analysis of the X-ray emission of Puppis A. Our data set constitutes by far the most sensitive single observation to date that captures the emission from the entire SNR. Furthermore, it offers relatively uniform exposure over its extent, eliminating the need to create mosaics from many individual observations.
Our paper is organized as follows:
Sect.~\ref{Data} presents basic characteristics of our data set and initial steps taken to ensure its correct treatment. We describe our methods and results in imaging and spectroscopic analysis in Sect.~\ref{Analysis}, with the core results of spatially resolved spectroscopy being presented in Sect.~\ref{Spectroscopy}. Finally, in Sects.~\ref{Discussion} and \ref{Summary}, we summarize our results and discuss their physical implications.

\section{Observations and data preparation\label{Data}}
The primary observation of Puppis A used in this work was carried out on 29/30 November 2019 as part of the calibration and performance verification campaign of eROSITA. Its main purpose was to calibrate the response and vignetting of the telescope.
The observation was performed in field-scan mode, covering a region of around $2\,\times\,2$ degrees, including the entire Puppis A SNR and a western region of Vela. The total duration of the observation was $60\,\si{ks}$. Due to telemetry constraints, an electronic ``chopper'' was set that discards every second frame taken by the cameras, yielding an effective observation duration of $30\,\si{ks}$. Furthermore, only the detectors with on-chip filters (TMs 1, 2, 3, 4, 6) were used. 
As the observation was performed in scanning mode, the effective spatial resolution of the data is similar to the survey average, at around $26\arcsec$ (half-energy width). 
Due to slightly varying scanning patterns, the observation was formally divided into two parts of approximately equal length (ObsIDs 700199 and 700200), separated by a gap of about $2.5 \,\si{ks}$.

For our entire analysis, we used the processing version {\tt c001} of the data set, identical to the data released as part of the eROSITA early Data Release (EDR).\footnote{See \url{https://erosita.mpe.mpg.de/edr/index.php}} The data were analyzed using the publicly released version of the eROSITA science analysis software, {\tt eSASSusers\_201009} \citep{Brunner21}.

As a first step, we used the task {\tt evtool} to merge the data from the individual TMs and the two sub-observations, allowing all valid patterns ({\tt pattern=15}), which left a total of around $39$ million valid events in the energy range $0.2-10.0 \,\si{keV}$.
The next step constitutes an important special treatment of the data, which was made necessary due to an error in the treatment of chopper values $>1$ in the {\tt c001} processing version of the data:\footnote{See also \url{https://erosita.mpe.mpg.de/edr/FAQ/}}
on one hand, the chopper settings were handled by introducing an individual good time interval (GTI) per valid frame (with a length around $50\,\si{ms}$). 
Therefore, in total, there are around $600\,000$ GTIs for each TM, which makes the runtime required for the computation of exposure-related quantities (exposure maps, ARFs) extremely long. 
In addition, a dead time correction factor (stored in the event file extensions {\tt DEADCOR}) was introduced to account for the exposure loss, which effectively divides the total exposure by the chopper value for a second, unnecessary time.  
To tackle both issues, we manually modified the {\tt GTI} extensions in the merged event file by joining all those GTIs constituting only of a single frame that were also separated only by the duration of a single frame. 
This step strongly speeds up downstream analysis, and corrects for the artificial exposure underestimation, as the factor $2$ is now incorporated in the {\tt DEADCOR} extensions only. 
We note that this issue is only present in the present processing version of the EDR data ({\tt c001}), and will eventually be corrected in a future public update of the data, making the outlined treatment obsolete. 

In addition to our main data set, we attempted to incorporate a second, more recent observation campaign of Puppis A \citep[see][]{Krivonos21}. This was carried out on 24/25 May 2021 as a grid of pointings toward the center and northeast of the SNR, with the purpose of cross-calibration with the Mikhail Pavlinsky ART-XC instrument \citep{Pavlinsky21}. 
Usable eROSITA data with a chopper-corrected total exposure around $10\,\si{ks}$ were obtained by the pointings with the eROSITA ObsIDs 730071$-$730092.\footnote{These observations are assigned the experiment numbers 12110052003$-$12110052044 in the SRG mission program, which are used to identify the corresponding ART-XC data set in \citet{Krivonos21}.} 
Unfortunately, less conservative instrument settings during these pointings lead to exceeded event quotas on board. Therefore, a significant fraction of recorded frames were lost, preventing the determination of accurate normalizations from extracted spectra.  
We therefore only use this data set for brief imaging analysis (see Sect.~\ref{Broadband}), as it nonetheless provides an improved spatial resolution in the northeast SNR quadrant due to the different pointing strategy than in the primary observation. 

\section{Analysis and results \label{Analysis}}
\subsection{Broad-band morphology \label{Broadband}}

\begin{figure}
\centering
\includegraphics[width=1.0\linewidth]{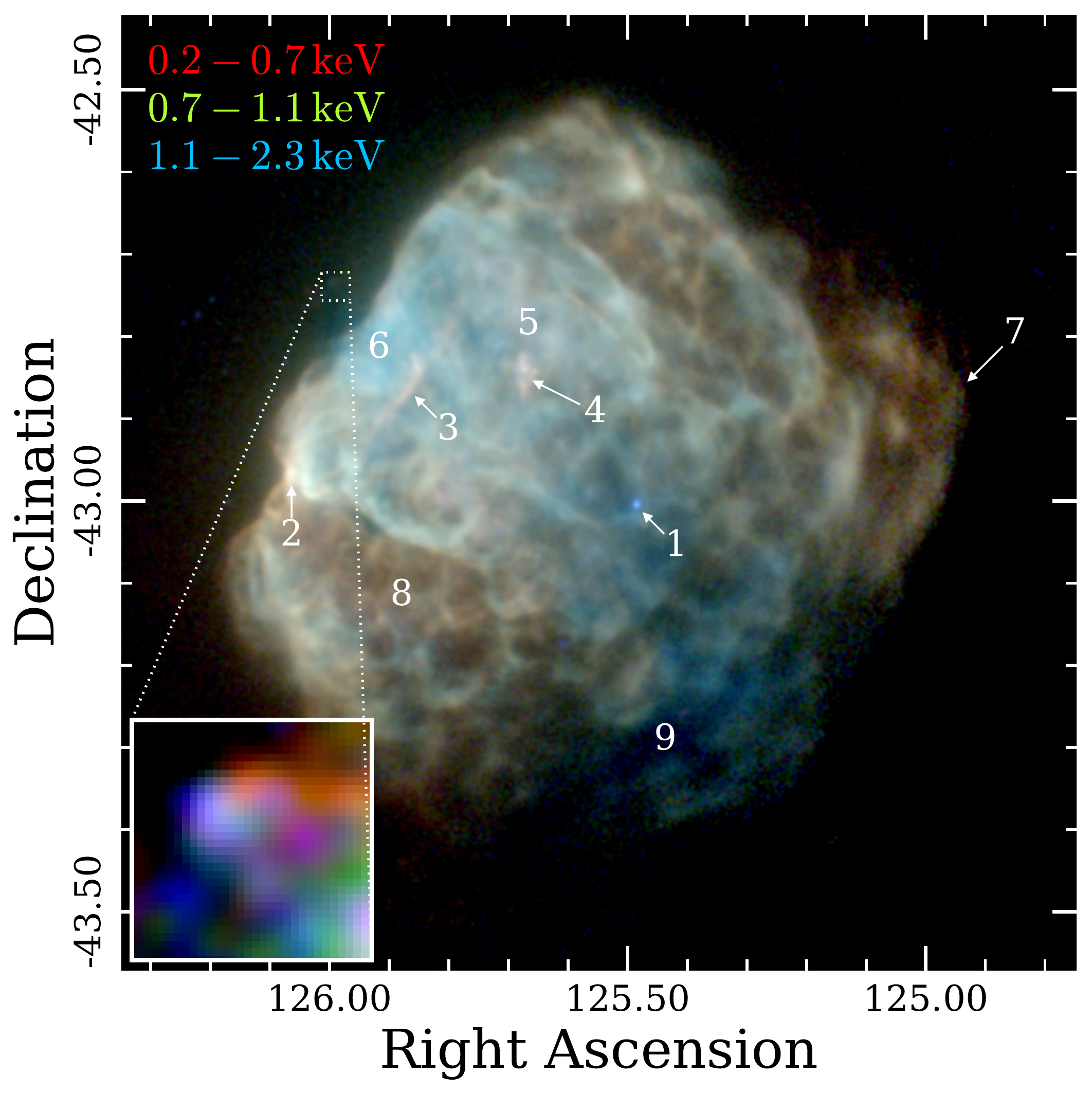} 
\caption{
Guide to the main features of Puppis A.
We show an exposure-corrected false-color image of Puppis A in the same energy bands as in Fig.~\ref{PuppisImage}, created from
the data set taken in May 2021, with an inset highlighting the location and shape of the bullet-like feature.  
The numbers label the following characteristic features of Puppis A discussed in the text:
(1) CCO, 
(2) BEK \citep[see][]{Hwang05}, 
(3) northeast filament \citep{Katsuda10}, 
(4) ejecta knot \citep{Katsuda08}, 
(5) ejecta-rich region \citep{Hwang08},
(6) northeast rim,
(7) western arc,
(8) faint southeast,
(9) southern hole \citep{Dubner13}}.
\label{Bullet}
\end{figure}

As a first visualization of the eROSITA view of the morphology of Puppis A, we created images and exposure maps in the broad energy band $0.2-2.3\,\si{keV}$, as well as in soft, medium and hard bands adapted to the SNR's spectrum at $0.2-0.7, \,0.7-1.1, \,1.1-2.3\,\si{keV}$, which closely match the bands of \citet{Dubner13}. 
We chose the $0.2-2.3\,\si{keV}$ range for our images, as it contains both the most sensitive region of the eROSITA response, and the vast majority of the X-ray emission of Puppis A. 
All images and exposure maps here and in the following were constructed using the {\tt evtool} and {\tt expmap} tasks, using an angular binning of $4\arcsec$. 
As the local background level, which below $2.3\,\si{keV}$ mainly originates from the nearby Vela SNR, is at least around an order of magnitude below the brightness level of Puppis A, except for its faintest and most absorbed regions, we did not attempt to subtract any background component to produce our broad-band images.  
We found that the noise in the images can be sufficiently suppressed already using a $\sigma = 6\arcsec$ smoothing kernel due to the excellent available statistics, with absolute count numbers ranging between around $\sim2\times10^{3} \, \si{ct.arcmin^{-2}}$ and $3\times10^{5} \, \si{ct.arcmin^{-2}}$ across the SNR. 

The resulting exposure-corrected broad-band and false-color images of Puppis A can be seen in Fig.~\ref{PuppisImage}. 
The global and small-scale morphology of Puppis A, first compiled in \citet{Dubner13}, is very well reproduced in our eROSITA observation. Numerous features, such as the presence of a characteristic strip of hard emission crossing the SNR from northeast to southwest, its hot CCO, an almost circular arc-like feature in the west, and numerous clumps and filaments across the SNR, are clearly visible. As far as the performance of eROSITA is concerned, it is particularly noteworthy that the image shown here was obtained using only a single observation with a total duration of $30\,\si{ks}$, distributed over a $2\times2$ degree field of view, leading to an effective vignetted exposure of Puppis A around $5\,\si{ks}$ in the relevant band.
In contrast, the image of \citet{Dubner13} consists of a mosaic of many {\it Chandra} and {\it XMM-Newton} observations, with a total of several hundreds of kiloseconds exposure time. 

Figure \ref{Bullet} displays 
an analogous false-color image to Fig.~\ref{PuppisImage},
created from the calibration observations carried out in May 2021. 
In this figure, we have labelled prominent features of Puppis A that are discussed here and in the following sections, in order to ease their identification by the reader.
While the displayed data set largely provides an identical impression to Fig.~\ref{PuppisImage}, it exhibits a somewhat better spatial resolution in the northeast of Puppis A, as the pointings were mainly aimed there. This reveals a previously unknown feature located outside the northeast rim of the remnant. Located at around $(\alpha, \delta) = (08^{\rm h}23^{\rm m}59^{\rm s}, -42^{\circ}44^{\prime})$, just outside the field of view of previous {\it Chandra} and {\it XMM-Newton} observations, it shows a cone-like morphology, and thereby resembles the well-known but much larger Vela ``shrapnels'', which are interpreted as fragments of ejecta outside the SNR shell \citep{Aschenbach95,Miyata01}. 
Unfortunately, the emission of the feature is comparatively faint, preventing further spectral analysis. Assuming a shocked plasma model typical for Puppis A (see Sect.~\ref{Spectroscopy}), the observed count rate of $0.05\,\si{ct.s^{-1}}$ suggests an incident flux of $\sim 4\times10^{-14}\,\si{erg.s^{-1}.cm^{-2}}$ in the $0.2-2.3\,\si{keV}$ band, on top of a dominant background.

\subsection{Narrow-band imaging and ratio images \label{Narrowband}}
\begin{figure}
\centering
\includegraphics[width=1.0\linewidth]{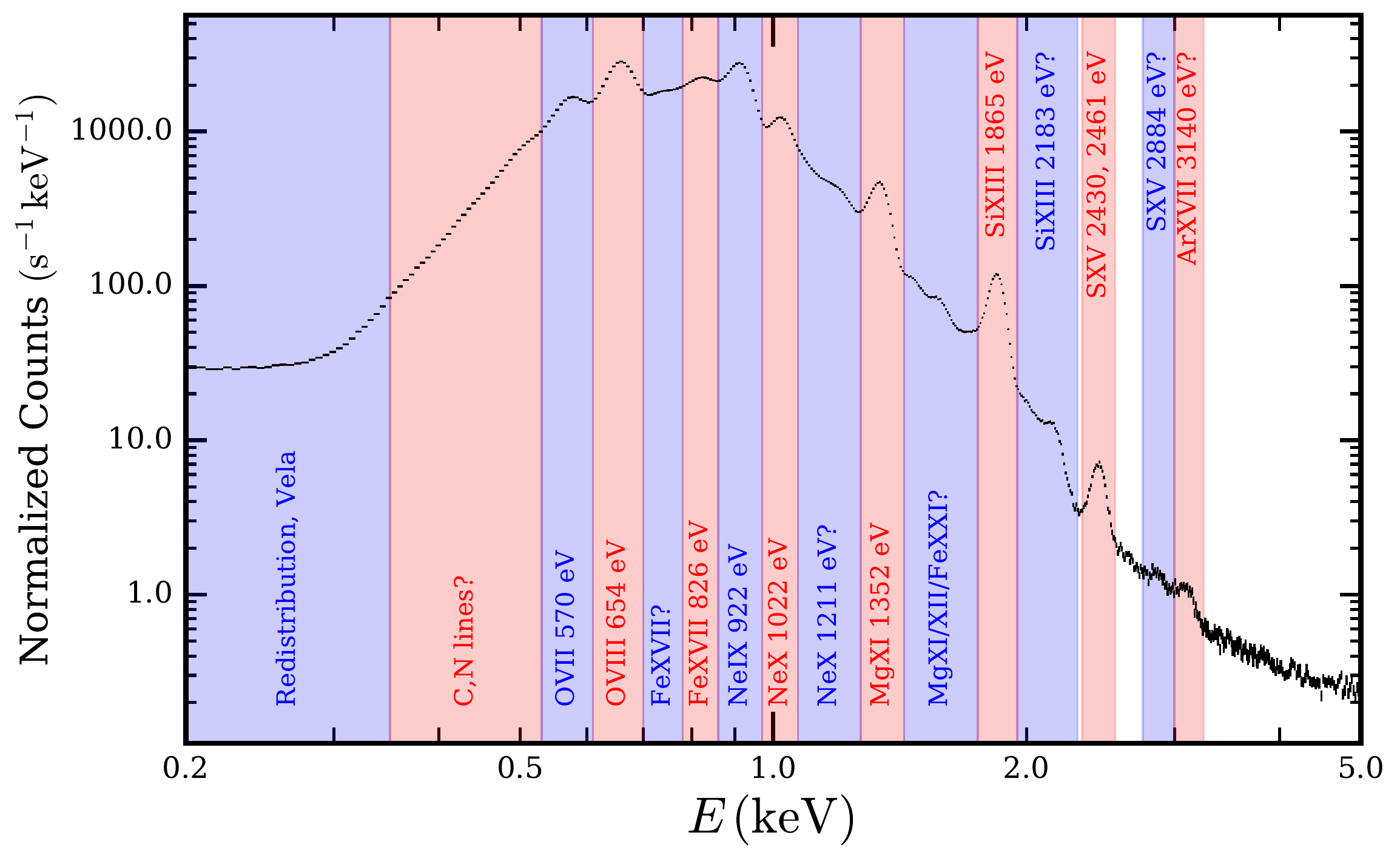} 
\caption{Integrated spectrum of the entire Puppis A SNR in the range $0.2-5.0\,\si{keV}$. The energy bands defined for narrow-band imaging are indicated in blue and red, together with the main contributions to the emission in the band.
Line components marked with a question mark are likely present but subdominant with respect to the continuum. }
\label{IntSpec}
\end{figure}

The improved spectral resolution and reduced low-energy redistribution of the eROSITA CCDs with respect to those of {\it XMM-Newton} EPIC-pn \citep{Meidinger21} allow us to obtain cleaner narrow-band images of Puppis A that isolate the emission of individual lines or line complexes. In order to define the ideal bands to isolate such lines, we extracted an integrated spectrum of the entire remnant using {\tt srctool}, and investigated the presence of strong lines in it (see Fig.~\ref{IntSpec}). While the physical meaningfulness of a detailed fit to this integrated spectrum would be limited due to the superposition of plasmas at different conditions and absorbed by different column densities, it is a useful tool to identify the presence also of weaker lines at high energies. 
Based on this integrated spectrum, we defined 16 narrow energy bands (indicated in Fig.~\ref{IntSpec}), covering strong and weak line complexes and pseudo-continuum regions across the energy range $0.2-3.25\,\si{keV}$. 

\begin{figure*}
\centering
\includegraphics[width=18.0cm]{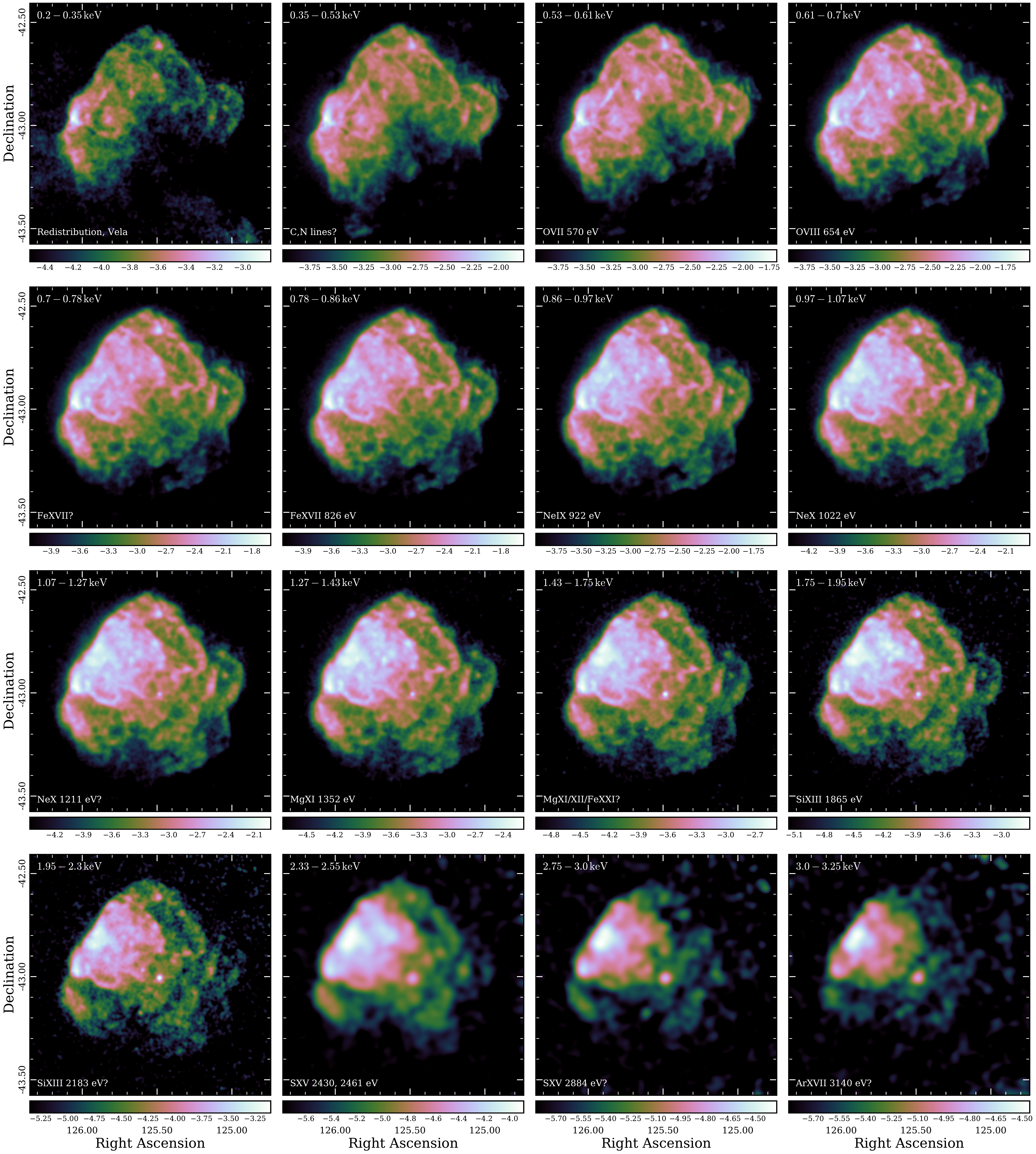} 
\caption{Exposure-corrected narrow-band images of Puppis A. 
For all bands below $2.3 \,\si{keV}$, we used a uniform smoothing kernel of $20\arcsec$, while for those above $2.3 \,\si{keV}$, we used $60\arcsec$. These kernels were chosen in order to ensure maximum comparability between bands, while taking into account the much poorer statistics at high energies.    
In each image, we have used a logarithmic color scale with the maximum at the peak count rate, and the minimum at the median count rate of all displayed pixels. 
The color bar  
underneath each panel indicates the displayed range of the logarithmic count rate per pixel, to illustrate the dynamic range of each image.
The upper left corner of each panel indicates the energy band covered by the image, while in the lower left corner, we indicate the physical content of the respective energy range. Components marked with a question mark are expected to be subdominant with respect to the (apparent) continuum.} 
\vspace{-5mm}
\label{NarrowBandImages}
\end{figure*}

An array of exposure-corrected images in these 16 bands is shown in Fig.~\ref{NarrowBandImages}. For each image, we used a logarithmic intensity scale whose zero point was set to the median brightness of all pixels, thus preserving comparability of the bands despite varying dynamic ranges. 
We chose to refrain from attempting to correct for the presence of background in this part of our analysis, for the following reasons: in the softest bands, the dominant background component, the Vela SNR, has a non-uniform and unknown morphology, making any attempted correction dependent on the assumption of a highly uncertain spatial template. In the hard bands ($> 2.3\,\si{keV}$), the dominant component is the instrumental background, which generally exhibits little flaring \citep{Freyberg20}, and is approximately uniform on the spatial scales of the instrumental field of view ($\sim 1^{\circ}$), which is why we consider its presence to have little impact on our imaging analysis.

\begin{figure*}
\centering
\includegraphics[width=18.0cm]{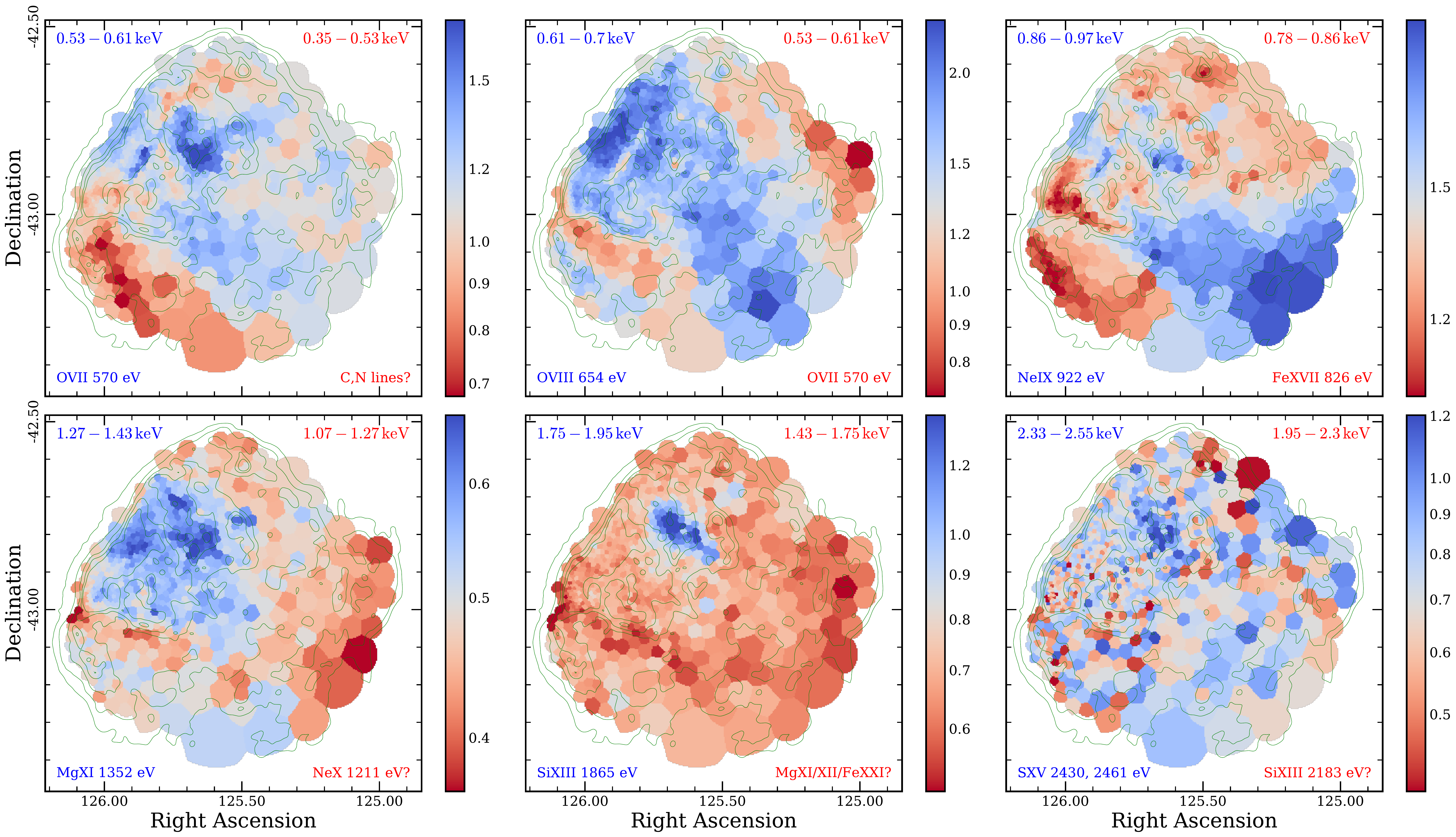} 
\caption{Maps of the ratio $H$ calculated from selected adjacent narrow energy bands. In each panel, we indicate larger relative strength of the harder (softer) band in blue (red), with the respective energy ranges indicated in the upper left (right) corner. 
The median $1\sigma$ uncertainties of $H$, which illustrate the typical statistical noise level in each panel, are (from left to right) $0.029$, $0.032$, $0.023$ in the top row, and $0.015$, $0.040$, $0.13$ in the bottom row.
The green contours correspond to the broad-band surface brightness of Puppis A, as indicated in Fig.~\ref{PuppisImage}.}
\label{DiffImages}
\end{figure*}

One of the most prominent characteristics visible is the strong absorption of the southwest quadrant of Puppis A, recognizable in the lack of emission at low energies. Furthermore, while at low energies, the BEK clearly dominates the SNR's emission, at the highest energies, its morphology is dominated by 
a region at the northeast rim,
and by the CCO, which emits as a hot blackbody.
More subtle differences between individual energy bands become visible upon close inspection, in particular if one overlays the individual narrow-band images.
\footnote{A movie looping through the emission of Puppis A in the individual energy bands shown in Fig.~\ref{NarrowBandImages} will be provided as supplementary material to the published version of this work.}

In order to make the presence of such differences between narrow-band images quantifiable and visible on paper, we created maps displaying the ratio of selected energy bands in the following way:
first, in order to obtain sufficient signal in all regions, we decided to rebin rather than smooth the data in each image, to preserve the independence of neighboring regions and not create any artifacts close to edges. 
To achieve this, we used the adaptive Voronoi binning algorithm by \citet{Vorbin}, and required a minimum signal-to-noise ratio in the $0.2-2.3\,\si{keV}$ band of $S/N=200$. 
Prior to the binning, the CCO was masked in the input image, as we are mostly interested in the diffuse emission in its surroundings.

In order to obtain diagnostic maps of plasma conditions and elemental abundances, we chose to compare selected adjacent energy bands, such that the relative difference introduced by spatially variable absorption columns is reduced as much as possible, and does not fully mask the effect of locally varying physical conditions.
To put these comparisons on a quantitative scale, we created maps of the band ratio $H$, which we defined as 
\begin{equation}
    H = \frac{R_2/A_2}{R_1/A_1},    
\end{equation}
where $R_1$ ($R_2$) refers to the vignetting-corrected count rate in the softer (harder) of the two adjacent bands, and $A_1$ ($A_2$) describes the on-axis effective area of the telescope averaged over the respective energy range. This effective area correction allows to directly compare observed line strengths, removing the effect of the telescope's energy-dependent response.

The resulting maps of the ratio $H$ between selected energy bands are shown in Fig.~\ref{DiffImages}.
Most panels in this figure 
reflect the strength of emission lines of metals typical for ejecta, as they display the ratio of a line-dominated to a \mbox{(pseudo-)continuum-dominated} band. While these band ratios are of course also influenced by other factors, such as plasma temperature or ionization age, strong line emission may point toward a large abundance of the respective element. 
For instance, a prominent peak of emission line strength from the species \ion{O}{vii}, \ion{Ne}{IX}, \ion{Mg}{XI}, \ion{Si}{XIII}, and \ion{S}{XV} is visible in a region located somewhat north of the SNR center, corresponding to the features labelled 4 and 5 in Fig.~\ref{Bullet}. This is consistent with the previously established presence of ejecta there \citep[see][]{Katsuda08,Katsuda13,Hwang08}. 
In contrast, for instance, the region of the BEK does not show any apparent enhancement in these emission lines, as expected from emission dominated by the collision of the blast wave with a density enhancement of the ISM \citep{Hwang05}. 

From a physical standpoint, the top central panel of Fig.~\ref{DiffImages} has a somewhat different interpretation: it displays the line ratio of two ionization states of oxygen, i.e. \ion{O}{viii} to \ion{O}{vii}. 
Consequently, a relative enhancement of the ratio indicates that the material exists in a higher ionization state compared to the average. Conversely, a relative deficiency corresponds to a lower ionization state, introduced either by colder, or by underionized, meaning recently shocked, plasma.
Comparing our results to the line ratio map published by \citet{Hwang08}, we recover the low degree of ionization of plasma in the northeast filament, as well as in 
a relatively faintly emitting region in the southeast (labelled as 8 in Fig.~\ref{Bullet}).
Similarly, we find a relative deficiency of \ion{O}{viii} emission along the entire northwest rim of Puppis A, which may be connected to a systematically smaller plasma temperature, particularly for the western arc, in accordance with its very soft broad-band emission (see Fig.~\ref{PuppisImage}). 
The northeast rim is the region most strongly dominated by \ion{O}{viii} emission. In combination with its increasingly dominant character toward high energies in imaging (see Fig.~\ref{NarrowBandImages}), this indicates that the northeast rim likely exhibits an elevated temperature and/or conditions close to collisional ionization equilibrium.

\subsection{Spatially resolved spectroscopy \label{Spectroscopy}}

\begin{figure*}[h!]
\centering
\includegraphics[width=18.0cm]{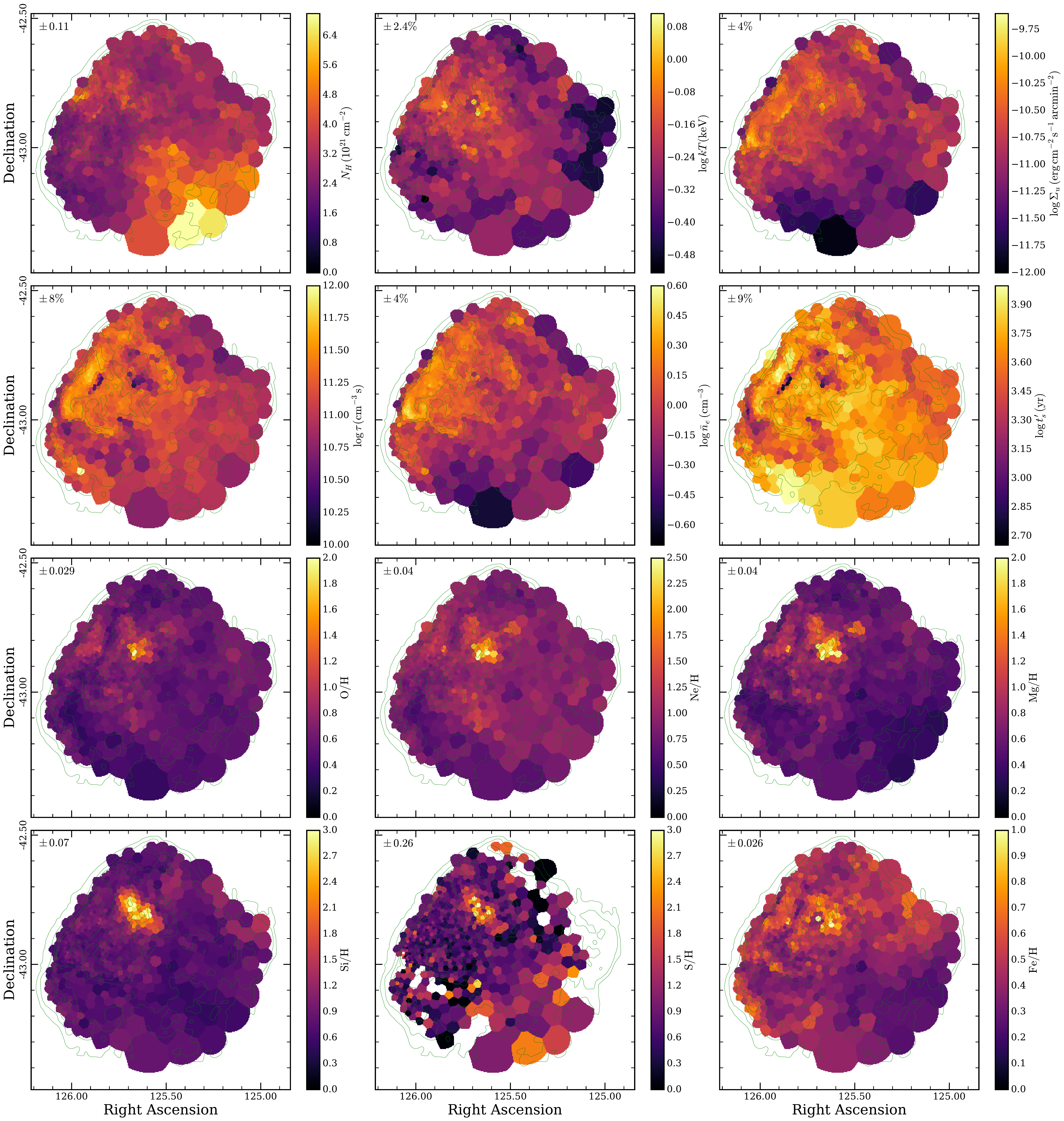} 
\caption{Parameter maps obtained from spectral fits to adaptively binned regions of $S/N = 200$. 
The color bar on the right of each panel indicates the displayed range of the respective parameter. 
The quantities displayed in the individual panels are the absorption column density $N_{\rm H}$, the plasma temperature $kT$, the unabsorbed surface brightness $\Sigma_{u}$ in the $0.2-5.0\,\si{keV}$ band, the ionization age $\tau$, the electron density proxy $\bar{n}_{e}$, the pseudo-shock-age $t^{\prime}_{\rm s}$, as well as the abundances of oxygen, neon, magnesium, silicon, sulfur, and iron normalized to solar values. 
In order to include only those regions which are dominated by emission from Puppis A, we masked all bins that extend up until the edge of our input image, as these are dominated by background.
In addition, for the sulfur abundances, regions with larger absolute errors than $0.8$ were masked. 
In the upper left corner of each panel, we indicate the typical statistical uncertainty of the displayed quantity, which corresponds to the median over all bins of the distribution of the $1\sigma$ error. 
For all logarithmically displayed quantities, the characteristic fractional error is reported in per cent. For those displayed on a linear scale, the error is given in the same units as indicated on the color bar.
The green contours in each panel correspond to the broad-band count rate of Puppis A, shown in Fig.~\ref{PuppisImage}.}
\label{TesselationImage}
\end{figure*}

In order to obtain a quantitative and in-depth picture of the physical conditions of the X-ray emitting plasma in Puppis A, it is necessary to go beyond the computation of band ratios, by forward-modelling the spectra extracted from different regions of the SNR. In order to achieve that, similarly to Sect.~\ref{Narrowband}, we used the adaptive Voronoi binning scheme of \citet{Vorbin} to define extraction regions from the broad-band ($0.2-2.3\,\si{keV}$) count image, masking the location of the CCO. The resulting bins were saved as bit masks in FITS files, and individually fed into the eSASS task {\tt srctool} to extract the corresponding spectra and ARFs.\footnote{Since {\tt srctool} currently writes only the default RMF file, irrespective of, e.g., detector coordinates, we only extracted a single RMF, which we used to fit all spectra.} 
In order to test different trade-offs between photon statistics and spatial resolution, we performed three separate runs, with target signal-to-noise ratios of $S/N = 200/ 300/ 500$, resulting in $772/345/114$ bins containing around $40\,000/90\,000/250\,000$ counts, respectively. 

We used PyXspec,\footnote{\url{https://heasarc.gsfc.nasa.gov/xanadu/xspec/python/html/index.html\#notes}} the Python implementation of Xspec \citep[version 12.11.0,][]{Arnaud96}, to fit the spectra from our individual regions. 
To describe the emission of Puppis A, we used a model of a plane-parallel shocked plasma \citep{Borkowski01} with non-equilibrium ionization (NEI), and foreground absorption following the T\"ubingen-Boulder model with the corresponding abundances \citep{Wilms00}. This model is expressed as {\tt TBabs*vpshock} in Xspec. 
Similarly to previous works, we thawed the abundances of the most prominent line-emitting elements in the spectral range of Puppis A (O, Ne, Mg, Si, S and Fe), while all remaining metal abundances were fixed to solar values. The lower limit of the ionization timescale as well as the redshift were fixed to zero.
Attempts to leave the redshift free to constrain line-of-sight velocities are not viable with the given data, as the absolute energy calibration in the {\tt c001} processing is not yet at the necessary level to allow for the reliable detection and quantification of the expected blue- or redshifts from velocities on the order of $\sim 1500\,\si{km.s^{-1}}$ \citep{Katsuda13}. 
While our model is not overly complex, we consider it a useful approximation to the observed spectra, which can be employed to constrain the average chemical and physical plasma properties throughout the SNR.

The background was modelled with three additional physical model components: 
first, the instrumental background is expected to be more or less homogeneous in spectral shape, irrespective of sky location. This can be justified with the rare occurrence of soft proton flares during solar minimum at the L2 environment of eROSITA and the weak nature of fluorescence lines on the detector \citep{Freyberg20}. Therefore, the only parameter left free in our fits was a multiplicative constant setting the global background normalization in the respective region, whereas its relative shape was taken from models fitted to filter-wheel-closed data.\footnote{Models are available at \url{https://erosita.mpe.mpg.de/edr/eROSITAObservations/EDRFWC}} 
The model used consists of a combination of two power laws and several Gaussian emission lines, reflecting the particle-induced background continuum and instrumental fluorescence lines, respectively. As these components do not correspond to actual X-ray photons, the instrumental background model was not multiplied by the ARF.

Second, we modeled the nonthermal X-ray background introduced by unresolved AGN using a single absorbed power law with a photon index of $1.46$ and fixed normalization per unit area determined by \citet{XRB}.

Finally, the thermal X-ray ``background'' in the relevant sky area requires careful treatment. Its dominant component at low energies is in fact a foreground, originating from the very extended Vela SNR, whose emission cannot be assumed to be spatially uniform. 
Vela is characterized by quite soft emission, generally at a lower surface brightness than Puppis A. However, we found that in particular for regions in which the observed emission of Puppis A is faint at low energies (e.g., in the south), its inclusion has a significant (improving) effect on the spectral fitting. 
We modelled the background contribution in the following way:
we extracted a spectrum from a nearby bright Vela filament within the field of view of our observation, and fitted it using a {\tt TBabs*(vapec+vpshock)} model \citep[e.g.,][]{Silich20}, in addition to the aforementioned instrumental and non-thermal components. 
Our primary goal in this step was finding a good fit to the data, rather than obtaining a straightforward physical interpretation for the background model. We therefore left the abundances of all relevant elements free to vary independently to achieve maximum model flexibility.
During the fit of the spectra of Puppis A, the shape of this model was fixed, and only the overall normalization of the thermal background left free. 
Note that Vela is located at a distance of approximately $290\,\si{pc}$ \citep{Dodson03}, around one quarter of the distance to Puppis A. This means that the soft portion of the thermal fore-/background component is subject to much less absorption than the emission of Puppis A, and any correlation between source and background absorption column densities will be rather weak. This justifies fixing the column density in our background template to a uniform value for all regions, rather than attempting to correct for its unknown possible variation across Puppis A.

Even though virtually all physical emission from Puppis A is encountered between around $0.4$ and $4.0\,\si{keV}$, in order to be able to robustly constrain the normalization of both the soft thermal X-ray background and the high-energy tail of the instrumental background, we used a very wide energy range of $0.20-8.50 \,\si{keV}$ for spectral modelling. In order to avoid rebinning the spectra prior to their analysis, we used Cash statistics for all our fits \citep{Cash}.
For the source spectra, elemental abundances were first kept frozen and then thawed after an initial run. Finally, the {\tt error} command was run to reduce the odds of converging toward secondary minima, and to obtain a rough estimate of the statistical uncertainty of our parameters.   
By repeating the outlined procedure for all adaptively binned spectra, and recombining the results with the celestial location of the regions, we created maps of the physical parameters constrained by our model. 

The Xspec expression for the emission measure, $\rm EM$, acts as a normalization of the model and encodes information about the density of the emitting plasma:\footnote{\url{https://heasarc.gsfc.nasa.gov/xanadu/xspec/manual/node213.html}} 
\begin{equation} \label{EMEquation}
    \mathrm{EM} = \frac{10^{-14}}{4\pi d^2}\int n_{e}n_{\rm H} \,\mathrm{d}V, 
\end{equation}
where $d$ describes the distance to Puppis A, $n_{e}$ and $n_{\rm H}$ describe the post-shock number densities of electrons and hydrogen atoms, respectively, and the integral runs over the emitting physical volume $\mathrm{d}V$. From this, using the fitted emission measure normalized by the angular size $\Omega$ of the extraction region (in steradian), one can derive the quantity:
\begin{equation} \label{ne}
    \bar{n}_{e} = 10^7 \sqrt{\frac{4\pi c_{e}}{D_{\rm LoS} f} \times \frac{\rm EM}{\Omega}},
\end{equation}
where $c_{e} \approx 1.2$ is equal to the ratio of electrons to hydrogen atoms for a typical ionized plasma with approximately solar abundances. 
Furthermore, $D_{\rm LoS}$ is an estimate of the depth of the emitting plasma along the line of sight, which we set to  $20\,\si{pc}$ here. This assumes a depth similar to the angular extent of Puppis A at a distance of $1.3\,\si{kpc}$ \citep{Reynoso17}.
Finally, the filling factor $f$ accounts for an inhomogeneous density distribution within the considered emitting region. For simplicity, at this point, we omit $f$, so that the resulting quantity is equivalent to $\bar{n}_{e}=\sqrt{\langle n_{e}^2 \rangle}$, where the average runs over the full volume of depth $D_{\rm LoS}$. The main advantage of our assumptions is that $\bar{n}_{e}$ can be calculated without invoking uncertain assumptions dependent on the three-dimensional density structure of Puppis A. Thus, while it constitutes an average over a large, likely inhomogeneous volume, $\bar{n}_{e}$ serves as a useful proxy to the density of the post-shock plasma.
The true average density over the full considered volume of an extraction region will always be smaller than $\bar{n}_{e}$, unless the material is distributed perfectly homogeneously within the entire volume. However, the peak density, which also presents the largest relative contribution to the overall emission, will always be larger. 
In order to obtain quantitative mass estimates (see Sect.~\ref{MassEstimates}), it is necessary to make more realistic assumptions on the density structure of Puppis A.

The ionization age is defined as $\tau=n_{e}\,t_{\rm s}$, where $t_{\rm s}$ is the time since the emitting material was first struck by the shock wave \citep{Borkowski01}. It describes the upper limit to observed ionization timescales in the plasma, which in turn quantify the degree of departure from collisional ionization equilibrium (CIE). 
In order to account for the dependence of $\tau$ on the post-shock density and attempt to reconstruct the propagation history of the blast wave, we define the shock pseudo-age as $t^{\prime}_{\rm s} = \tau/\bar{n}_{e}$. This quantity should not be seen as an exact quantitative estimate of the collision time of the supernova shock wave with circumstellar material, as that would assume a truly uniform density distribution over the emitting volume as well as constant densities over time in the post-shock region. 
However, the measured distribution of $t^{\prime}_{\rm s}$ over the extent of Puppis A serves as an indicator of the relative recency of the interaction between shock and ISM (or ejecta) in different parts of Puppis A.

In the resulting parameter maps, we generally found that the trade-off between statistical noise and spatial resolution appeared optimal for the lowest binning threshold $S/N = 200$. In the following, we therefore discuss the parameter maps binned to $S/N=200$, which are displayed in Fig.~\ref{TesselationImage}.
However, for weakly constrained parameters (e.g., sulfur abundance), the coarser binning tends to lead to a clearer picture due to the improved suppression of noise. 
We therefore show similar plots with $S/N=300$ and $500$ in Figs.~\ref{TesselationImage_SNR300} and \ref{TesselationImage_SNR500} in the appendix.

\begin{figure*}[h!]
\centering
\includegraphics[width=18.0cm]{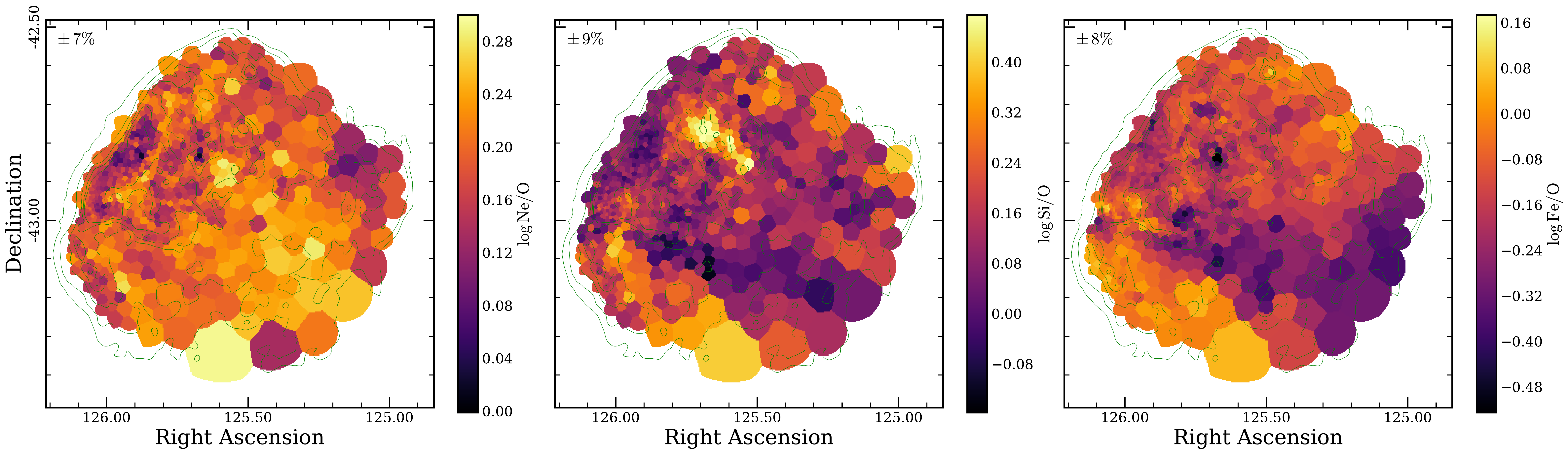} 
\caption{Ratio of selected elemental abundances with respect to solar, obtained from spectral fits to adaptively binned regions of $S/N = 200$ (see Fig.~\ref{TesselationImage}). 
The contours, region masking, and indication of the fractional uncertainty are as in Fig.~\ref{TesselationImage}. 
}
\label{MetalRatioImage}
\end{figure*}

\subsubsection{Absorption and plasma conditions}
We clearly find a varying degree of foreground absorption across the remnant, as previously suspected by numerous authors \citep[e.g.,][]{Aschenbach93,HuiBecker06,Dubner13}. 
While the distribution of measured hydrogen column densities seems to exhibit a relatively sharp lower cutoff around $N_{\rm H} \approx 2\times10^{21}\,\si{cm^{-2}}$ (see also Fig.~\ref{PlasmaCorrels}), the southwest quadrant of Puppis A appears to be strongly absorbed with measurements up to around $7\times10^{21}\,\si{cm^{-2}}$, consistent with the relatively hard emission there (see Figs.~\ref{PuppisImage} and \ref{NarrowBandImages}) and with the detection of cold foreground dust in this region  \citep{Arendt10,Dubner13}. 
The region in which the strongest far-infrared emission at $160\,\si{\mu m}$ is found is consistent with what appears like a ``hole'' in the soft X-ray emission toward the southern edge of the shell (see Fig.~\ref{Bullet}). However, due to insufficient photon statistics in this area, in our $N_{\rm H}$ map, this region is mixed with neighboring, less absorbed regions, making it challenging to infer what the true maximum column density toward Puppis A is. Therefore, we dedicate some further attention to this region in Sect.~\ref{DetailedFitSection}. 

Throughout the SNR, there are a few subtle features of enhanced absorption, which appear significant considering that statistical errors are only estimated to be on the order of $1\times10^{20}\,\si{cm^{-2}}$:
at the very northeast rim of the SNR, there appears to be a small-scale enhancement of absorption with measured column densities up to $5\times10^{21}\,\si{cm^{-2}}$, coincident with an apparent indentation of the rim. An explanation for this could be additional absorption introduced by a possibly cospatial density enhancement of the ISM toward the Galactic plane, 
visible also at infrared wavelengths \citep{Reynoso17,Dubner13,Arendt10}, with which the supernova shock wave is colliding. 
Furthermore, we note that the northeast filament 
also appears to show enhanced absorption (see Sect.~\ref{DetailedFitSection}). 

Similarly to the broad-band count rate, our map of absorption-corrected surface brightness $\Sigma_{u}$ exhibits a large gradient, ranging over about two orders of magnitude across the SNR. The highest fluxes are found at the northeast rim and the BEK, 
with values up to $\Sigma_{u} \sim 10^{-10} \,\si{erg.s^{-1}.cm^{-2}.arcmin^{-2}}$.
By integrating the flux of the source model over all unmasked regions in Fig.~\ref{TesselationImage}, we estimate the total unabsorbed flux of Puppis A to be 
$F_{u} = 2.64\times10^{-8}\,\si{erg.s^{-1}.cm^{-2}}$ in the $0.2-5.0\,\si{keV}$ band, which corresponds to an intrinsic luminosity $L_{x} = 5.3\times10^{36} \,\si{erg.s^{-1}}$ 
at a distance of $1.3\,\si{kpc}$ \citep{Reynoso17}. 
The dominant uncertainty in this measurement is not of statistical nature, as the statistical error of the flux, derived by adding the uncertainties of all bins in quadrature, is only at a level of $0.2\%$. Instead, it is caused by the combined systematic effects of effective area calibration uncertainties (see also Sect.~\ref{CCOSpec}), the particular model choice, and the somewhat arbitrary definition of the SNR extent. We estimate the combined systematic uncertainty of our measurement to be on the order of $15\%$, if and only if one assumes the used spectral model to be the correct one. If one wanted to include the possibility that, for instance, a two-component model may be closer to reality in some regions, larger uncertainties would likely apply. 
Converting to the $0.3-8.0\,\si{keV}$ energy range, the corresponding flux of $F_{u} = 2.43\times10^{-8}\,\si{erg.s^{-1}.cm^{-2}}$ (implying $L_{x} = 4.9\times10^{36} \,\si{erg.s^{-1}}$) is consistent with the broad uncertainty range derived by \citet{Dubner13}. However, it appears discrepant with the measurement of $F_{u} = (1.5\pm0.2)\times10^{-8}\,\si{erg.s^{-1}.cm^{-2}}$ for the same energy range by \citet{Silich20}. This is likely primarily caused by the difference between a spatially resolved and a spatially integrated treatment of the emission. We find a variation of the absorbing column density by a factor of a few over the SNR, whereas \citet{Silich20} find weaker overall absorption with a column density of $N_{\rm H} \sim 2\times 10^{21}\,\si{cm^{-2}}$, which would imply a larger fraction of intrinsic SNR flux reaching the observer. Given the smoothly varying structure of our $N_{\rm H}$ map, we believe that our flux measurement is more reliable. The resulting luminosity, and thereby the energetics of Puppis A, is subject to the additional systematic error of the assumed distance to Puppis A \citep{Reynoso17}, in addition to the factors affecting the flux measurement.

The line-of-sight density average of the post-shock plasma, $\bar{n}_{e}$, is rather tightly correlated with $\Sigma_{u}$, as it is computed based on the normalization of the spectrum. Unsurprisingly, the highest densities are found along the northeast rim and at the BEK, where it has been established that the shock wave is interacting with a small-scale density enhancement, possibly a molecular cloud \citep{Hwang05}. Densities in the south and west of the SNR are found to be up to an order of magnitude lower, consistent with a thinner environment in the direction away from the Galactic plane.

The distribution of the plasma temperature $kT$ over the SNR exhibits a median (and $68\%$ central interval) of $kT = 0.55_{-0.09}^{+0.14}\,\si{keV}$ (see Fig.~\ref{PlasmaCorrels}).
A few regions appear noteworthy in our map of $kT$. For instance, we find localized very high plasma temperatures close to the northeast filament and the ejecta knot \citep[see][]{Katsuda08,Katsuda10} at around $1.0\,\si{keV}$. This should however be treated as a somewhat uncertain finding, as regions rich in ejecta would realistically require modelling with multiple emission components (see Sect.~\ref{Modelchoice}), which may affect the inferred plasma temperature. 
In contrast, the western arc of Puppis A is found to exhibit temperatures as low as $0.35\,\si{keV}$. It may be possible that 
the soft emission here could be related to the existence of a local ISM cavity that would explain the almost circular structure of this feature.
An extended region along the northeast rim stands out quite clearly, with coherently elevated temperatures spanning the range $0.7-0.8\,\si{keV}$, which is a highly significant enhancement considering the typical statistical noise level of only $2.4\%$. Together with comparatively large ionization ages on the order $\tau\sim3\times10^{11} \,\si{cm^{-3}.s}$ measured here, the high plasma temperature provides a convincing explanation for the brightness of this region above $2 \,\si{keV}$.

Finally, a fascinating picture is offered by the map of the shock pseudo-age $t^{\prime}_{\rm s}$: numerous regions of freshly shocked plasma appear quite prominent. First, the northeast filament shows by far the lowest values of $t^{\prime}_{\rm s}$,
making this the region which appears to have experienced the most recent shock interaction in Puppis A. Similarly, the ejecta knot and part of the extended ejecta-rich region in the north of the SNR 
also appear to have been shocked quite recently. 
It should be noted that, since these features most likely do not extend along the whole line-of-sight through Puppis A, the inferred value of $\bar{n}_{e}$ is likely an underestimate of their true density, and consequently their true shock age is likely even lower than indicated in our map (see Sect.~\ref{DetailedFitSection}).
On larger scales, we recover the underionized nature of  relatively faintly emitting material in the southeast of Puppis A (feature 8 in Fig.~\ref{Bullet}) observed by \citet{Hwang08}.

Since most regions hosting a young shock are not located at the apparent edge of the remnant, a possible conclusion could be that their emission originates primarily behind or in front of the surrounding emission. Their apparent location on the ``inside'' of the shell would then be due to projection effects along the line of sight only. An alternative, which may be more likely for regions associated with ejecta, could be that the emitting material has recently interacted with
the reverse shock, reheating it and placing it in a state far from CIE. 
The western half of Puppis A does not appear to exhibit any similar recently shocked regions. On one hand, this may simply be an artifact of the fainter emission there, leading to larger bins for spectral extraction, which may mask the presence of small-scale features far from CIE.
On the other hand, one could also imagine that our inability to observe clear signatures of a reverse shock there \citep[see also][]{Katsuda10} 
may be due to a smaller amount of heated ejecta. Since the ISM is much thinner in the west, the mass swept up by the forward shock there is likely lower, leading to a less deep penetration of the reverse shock into the ejecta.

At the BEK, we find a quite clear dependence of $t^{\prime}_{\rm s}$ on the (angular) distance from the SNR center, with the outermost regions exhibiting the youngest shock.  
This illustrates very nicely the gradual penetration of the shock into the ISM, as regions further inside have naturally been struck by the shock at an earlier time. 
This is very much consistent with the results of \citet{Hwang05}, as they conclude that the ``bar'' and ``cap'' structures are the remnants of a mature shock-cloud interaction, which occurred $2000-4000\,\si{yr}$ ago. In contrast, the compact knot located at the easternmost edge of the BEK region is undergoing an intense interaction with dense ISM at the present time.  

A final interesting note to make is that the median value of $t^{\prime}_{\rm s}$ over all regions is around $4200\,\si{yr}$ (see Fig.~\ref{PlasmaCorrels}), which is comparable to quite precise kinematic age estimates of Puppis A \citep{Winkler88,Mayer20}, which range between $3700$ and $4600\,\si{yr}$. However, most likely, this apparent agreement is partly coincidental, as the assumptions going into the computation of $t^{\prime}_{\rm s}$ are extremely crude. For instance, the neglect of temporal changes in the density of the shocked plasma or of a possible non-uniform distribution of the emitting material likely would not hold under realistic conditions.

\subsubsection{Elemental abundances \label{Abundances}}
Our approach of decomposing the emission of Puppis A into many individual regions allows us to construct flux-weighted distribution functions of chemical abundances (see Fig.~\ref{AbundanceCorrels} in the Appendix). We find that the median metal abundances across the SNR, normalized by the abundance table of \citet{Wilms00}, are given by $\mathrm{O/H} = 0.62^{+0.15}_{-0.10}$, $\mathrm{Ne/H} = 0.94^{+0.20}_{-0.13}$, $\mathrm{Mg/H} = 0.60^{+0.22}_{-0.13}$, $\mathrm{Si/H} = 0.83^{+0.20}_{-0.13}$, $\mathrm{S/H} = 0.88^{+0.59}_{-0.29}$, and $\mathrm{Fe/H} = 0.45^{+0.12}_{-0.11}$ (errors marking the $68\%$ central intervals). The low median iron abundance and its relatively homogeneous distribution over the SNR do not indicate any significant enrichment of the emitting plasma with iron ejecta. 
Interestingly, the highest abundances relative to solar values across the remnant are found for neon, whereas the median oxygen abundance is considerably smaller (see Sect.~\ref{DiscEjDist}). 

Our abundance maps very clearly confirm the 
high metal content of the ejecta enhancements discussed in \citet{Katsuda08} and \citet{Hwang08} and labelled 4 and 5 in Fig.~\ref{Bullet}, as we find coherently elevated abundances of oxygen, neon, magnesium, silicon and sulfur there. Iron abundances appear to be only weakly (if at all) enhanced in this area compared to surrounding regions, as even the maximum measured abundances are only about solar.
Interestingly, while the peak of the light-element (O, Ne, and Mg) distribution is concentrated on the clumpy ejecta knot in the south of the enriched region, the heavier elements (Si and S) show enhancements spread over a larger, elongated region. 
This is illustrated more clearly by the maps in Fig.~\ref{MetalRatioImage}, which display the ratios $\mathrm{Ne/O}$, $\mathrm{Si/O}$, $\mathrm{Fe/O}$ corresponding to the abundance maps in Fig.~\ref{TesselationImage}. A clear peak in the silicon-to-oxygen ratio is visible for the extended ejecta-rich region, whereas the compact ejecta knot to its south shows a relative enhancement in oxygen abundance. 

The metal distribution across the remainder of Puppis A shows much more modest variations, which only appear visible for the lighter elements. For instance, there seems to be a weak enhancement in neon and magnesium at a location about $10\arcmin$ northwest of the prominent ejecta knot, coincident with a clumpy feature in the broad-band image of Puppis A.   
A further example is the hot northeast rim, which seems to show a relative oxygen enhancement, particularly visible in the $\mathrm{Ne/O}$ map. 
Generally, the maps displayed in Fig.~\ref{MetalRatioImage} appear to show some interesting large-scale structure, as for instance, a broad strip along the southeast rim of Puppis A appears to show 
enhanced $\mathrm{Si/O}$ and $\mathrm{Fe/O}$ ratios,
whereas the region to the north and west of it seems to be more enriched in oxygen in a relative sense. 

The distribution of sulfur abundance suffers from a large fraction of spurious measurements, with a significant number of unrealistically high ($\mathrm{S/H} \gtrsim3$) or low values.
We found that the majority of spurious elevated abundances can be suppressed by masking all those bins with a statistical $1\sigma$-error larger than $0.8$, which produces the map shown in Fig.~\ref{TesselationImage}. 
While the resulting map displays a convincing agreement between enhancements of sulfur and silicon in the extended ejecta-rich region, the entire western arc of Puppis A was masked as it yielded unreasonably high values for $\mathrm{S/H}$. 
A likely explanation for such behavior in the west of the SNR is the presence of a subdominant second emission component with higher plasma temperature. This component would produce much stronger \ion{S}{xv} line emission than expected from the best-fit single-temperature model with ISM abundances, thus driving the inferred sulfur abundance to unrealistically high, albeit formally statistically significant, values (see also region I in Sect.~\ref{DetailedFitSection}). 
Apart from the occurrence of systematic outliers, large statistical fluctuations, including the presence of bins with $\mathrm{S/H} \sim 0$, are apparent in the map. These are likely caused by the poor statistics of our spectra in the band most sensitive to emission from sulfur, which is above $2.3 \,\si{keV}$, where the eROSITA response exhibits a sharp drop. Indeed, the measured statistical errors are several times larger than for all our other abundance measurements, at a median $1\sigma$-error of $0.26$.

\subsubsection{Model choice \label{Modelchoice}}
\begin{figure}
\centering
\includegraphics[width=0.8\linewidth]{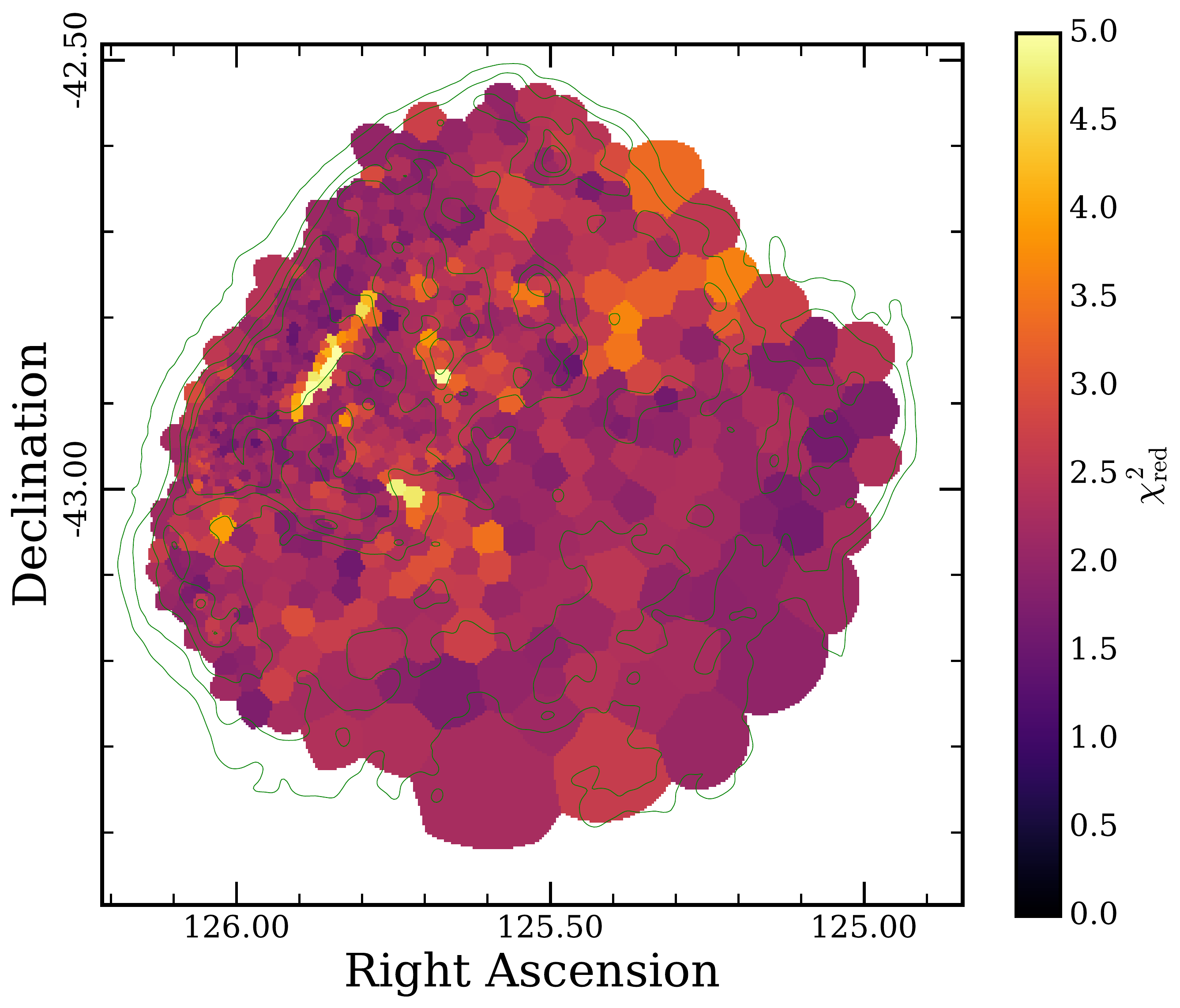} 
\caption{Map of the ``reduced'' $\chi^2$ statistic computed for the best-fit (parameters as displayed in Fig.~\ref{TesselationImage}) in each bin.
The number of degrees of freedom per bin varies between around 120 and 160 over the extent of the map, as each spectrum is rebinned individually to the target signal-to-noise ratio. }
\label{GoFImage}
\end{figure}

At this point, a valid question to ask may be whether the employed physical model is really a sufficient representation of the conditions in Puppis A. In order to investigate this matter, after rebinning the spectrum and best-fit model to a signal-to-noise ratio of five, we manually computed the ``reduced'' $\chi^2$ statistic for each spectrum, as a rough measure of goodness of fit. The resulting map is shown in Fig.~\ref{GoFImage}. It shows that, while the fits in a few regions are clearly suboptimal, the large majority of regions exhibit $\chi^2_{\rm red} \sim 2$. Considering the large number of total counts per spectrum and the early stage of energy and response calibration, we believe that this value indicates an acceptable fit.

A possible origin of the poor fit quality of our model in a few regions may be the assumption of 
only one emission component with a single set of parameters, which neglects the possibility of physically different conditions projected along the same line of sight. 
By evaluating the improvement of the fit likelihood when including a second {\tt vpshock} component, we verified that a second plasma component is indeed needed in those regions with a bad fit, in particular around the northeast filament. 
To tackle the observed issue there, we apply a more sophisticated treatment of the local ``background'' introduced by the emission of surrounding shocked ISM in Sect.~\ref{DetailedFitSection}.
In contrast to the northeast filament, those regions with $\chi^2_{\rm red} \sim 2$ in Fig.~\ref{GoFImage} generally show only minor improvements, indicating that the remaining imperfections of the fit there may indeed be mostly due to calibration uncertainties, not modelling issues. 
Generally, it should be noted that, while a two-component model naturally always leads to an equally good or improved fit, with the available spectral resolution, it is difficult to precisely constrain the individual parameters of both plasma components without making additional restricting assumptions, due to the large amount of parameter degeneracies introduced. 

As a further test to the validity of our particular model choice,  
we repeated our fitting procedure using the frequently employed single-ionization-timescale NEI plasma model {\tt vnei}. This approach qualitatively preserved many characteristic structures in our parameter maps, for instance the recently shocked regions and the abundance peaks in the ejecta regions. However, compared to the {\tt vpshock} model, it lead to deteriorated fit statistics in the vast majority of bins, the resulting maps appeared much noisier overall, and the fraction of outliers for weakly constrained parameters was larger. 
Furthermore, from a physical standpoint, the model of a plane-parallel shock plasma is likely more appropriate for realistic SNRs dominated by ISM interaction, as it assumes a continuous distribution of ionization timescales within the emitting plasma, rather than all material having been struck by the shockwave at a single point in time \citep{Borkowski01}.
For the given data set, we therefore believe that an absorbed single-component {\tt vpshock} model constitutes the optimal compromise between interpretability and physical accuracy of our spectral model.

\subsection{Detailed modelling of selected regions \label{DetailedFitSection}}

\begin{figure}
\centering
\includegraphics[width=1.0\linewidth]{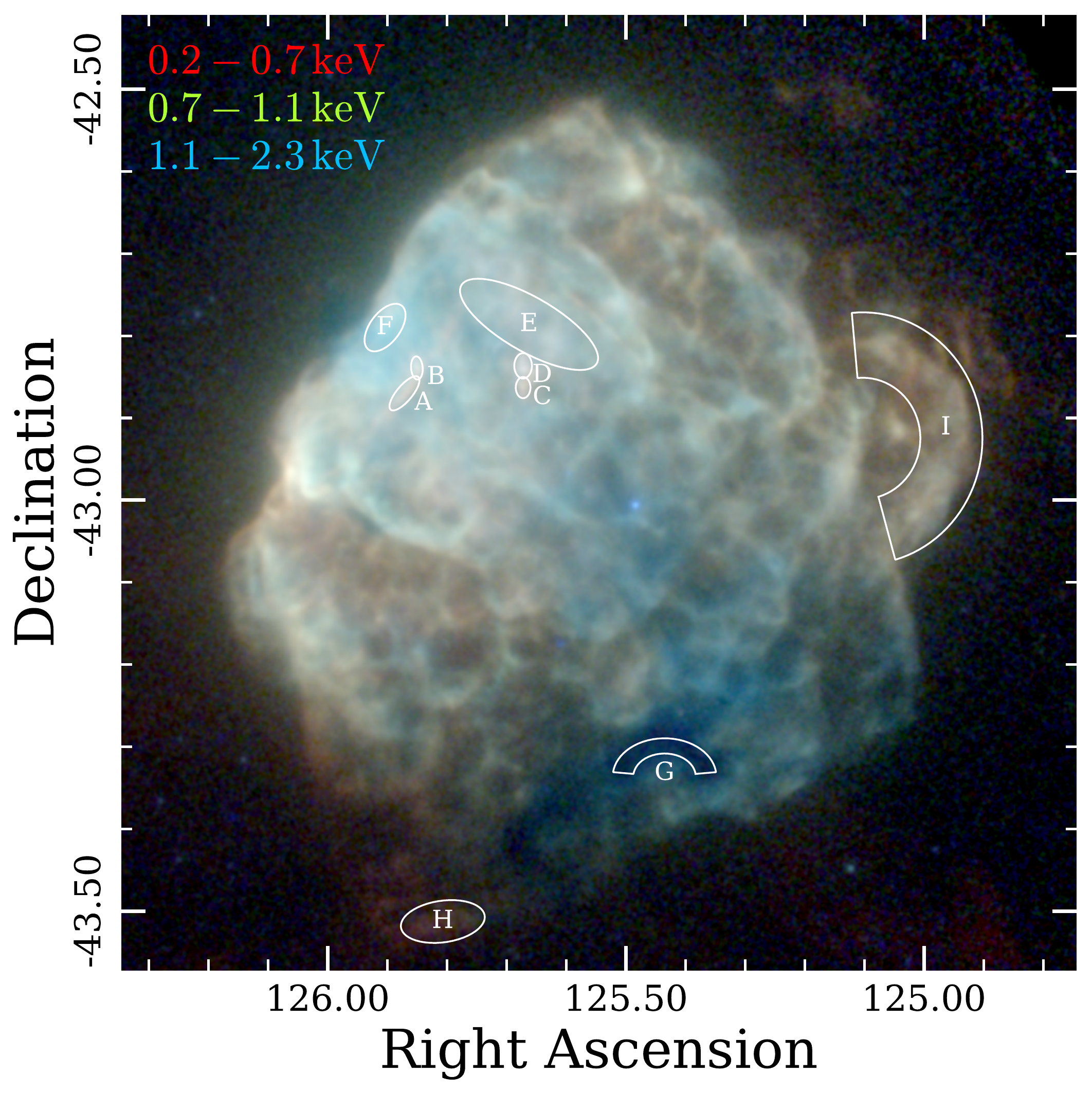} 
\caption{Regions for the extraction of the spectra in Fig.~\ref{DetailedFits} overlaid on the false-color image of Puppis A (same data as in Fig.~\ref{PuppisImage}). To highlight the faint feature H, the zero point of the color scale was set lower than in Fig.~\ref{PuppisImage}. 
}
\label{DetailedFitRegions}
\end{figure}

\begin{figure*}[h!]
\centering
\includegraphics[width=6.0cm]{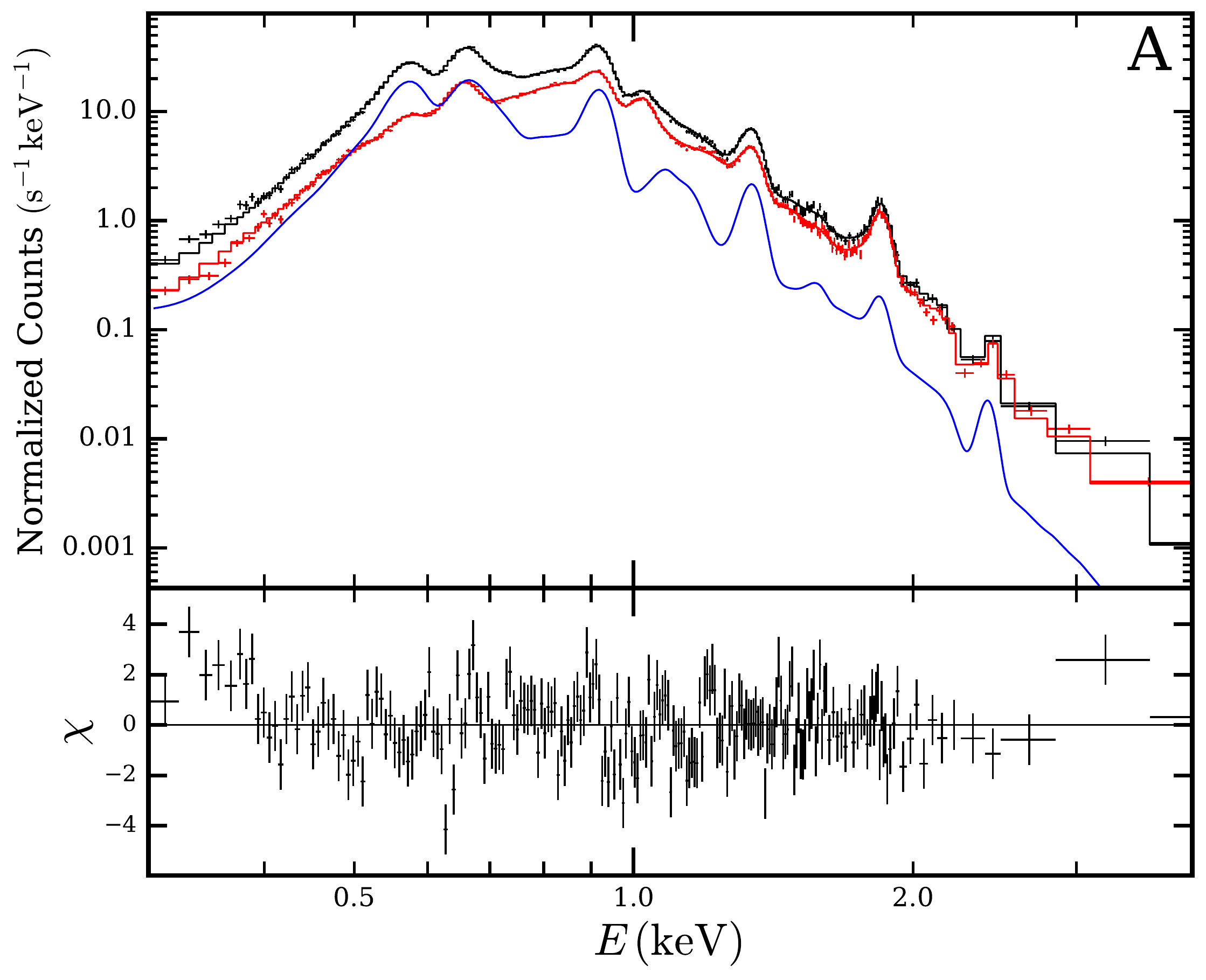} 
\includegraphics[width=6.0cm]{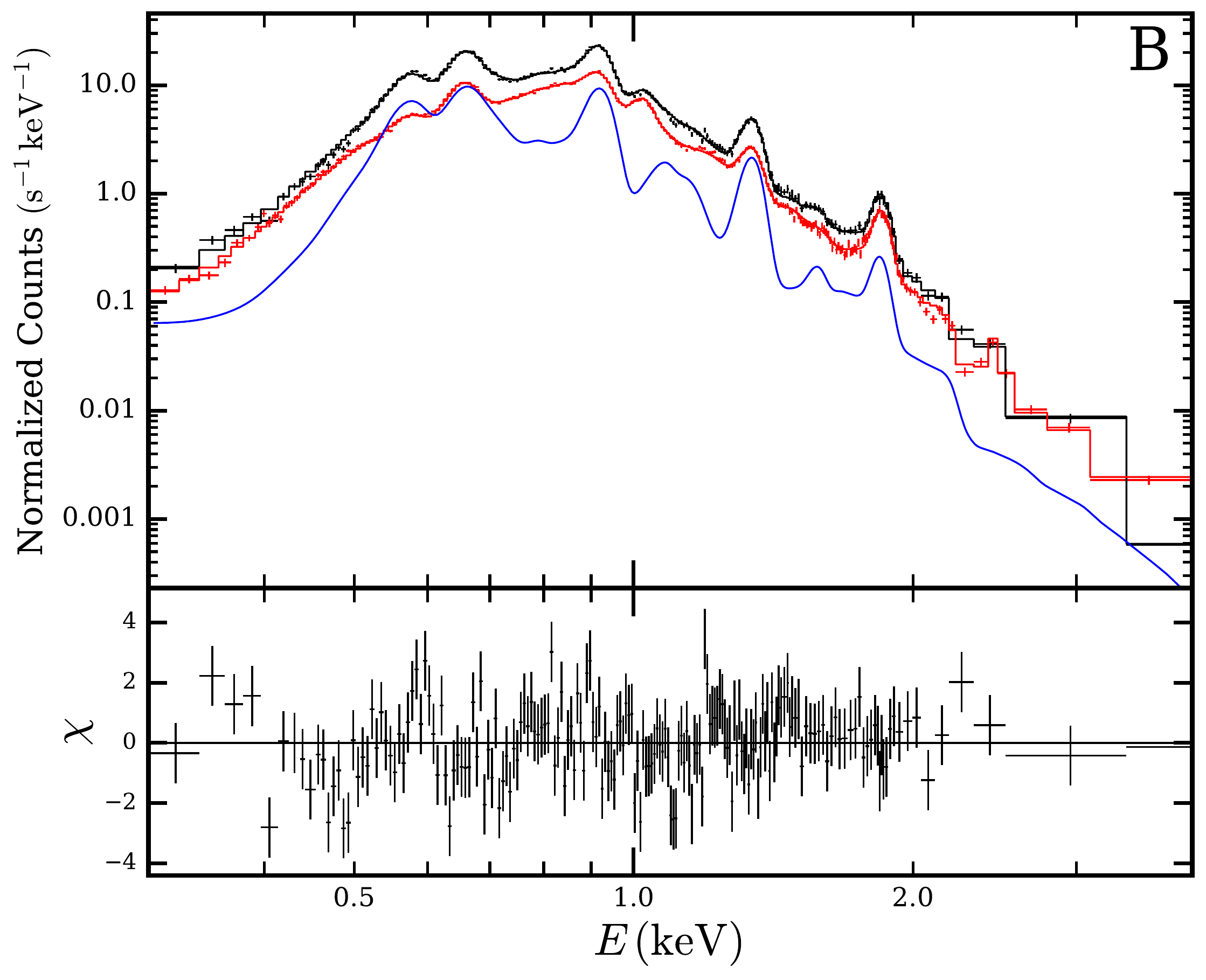}   
\includegraphics[width=6.0cm]{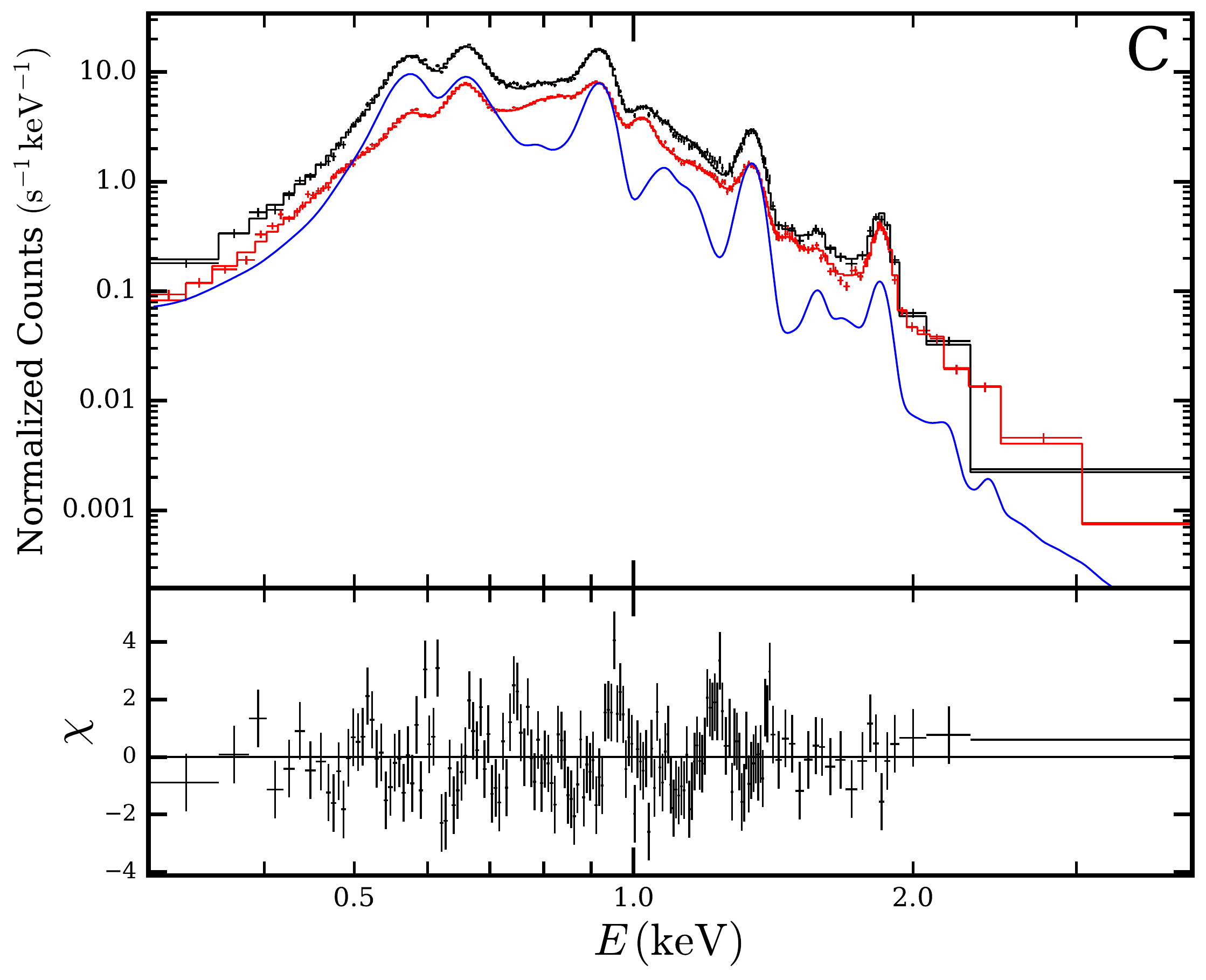} \\
\includegraphics[width=6.0cm]{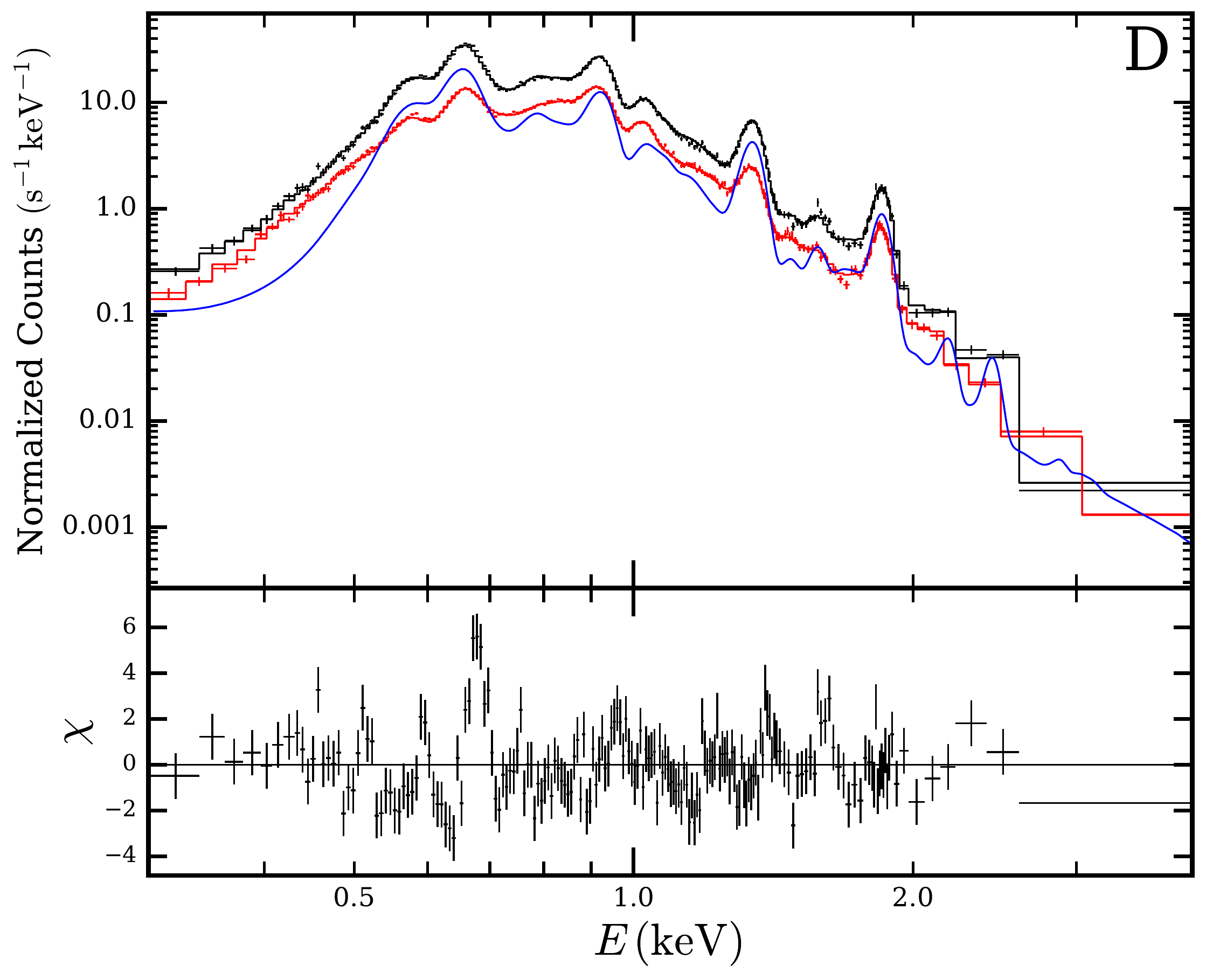}   
\includegraphics[width=6.0cm]{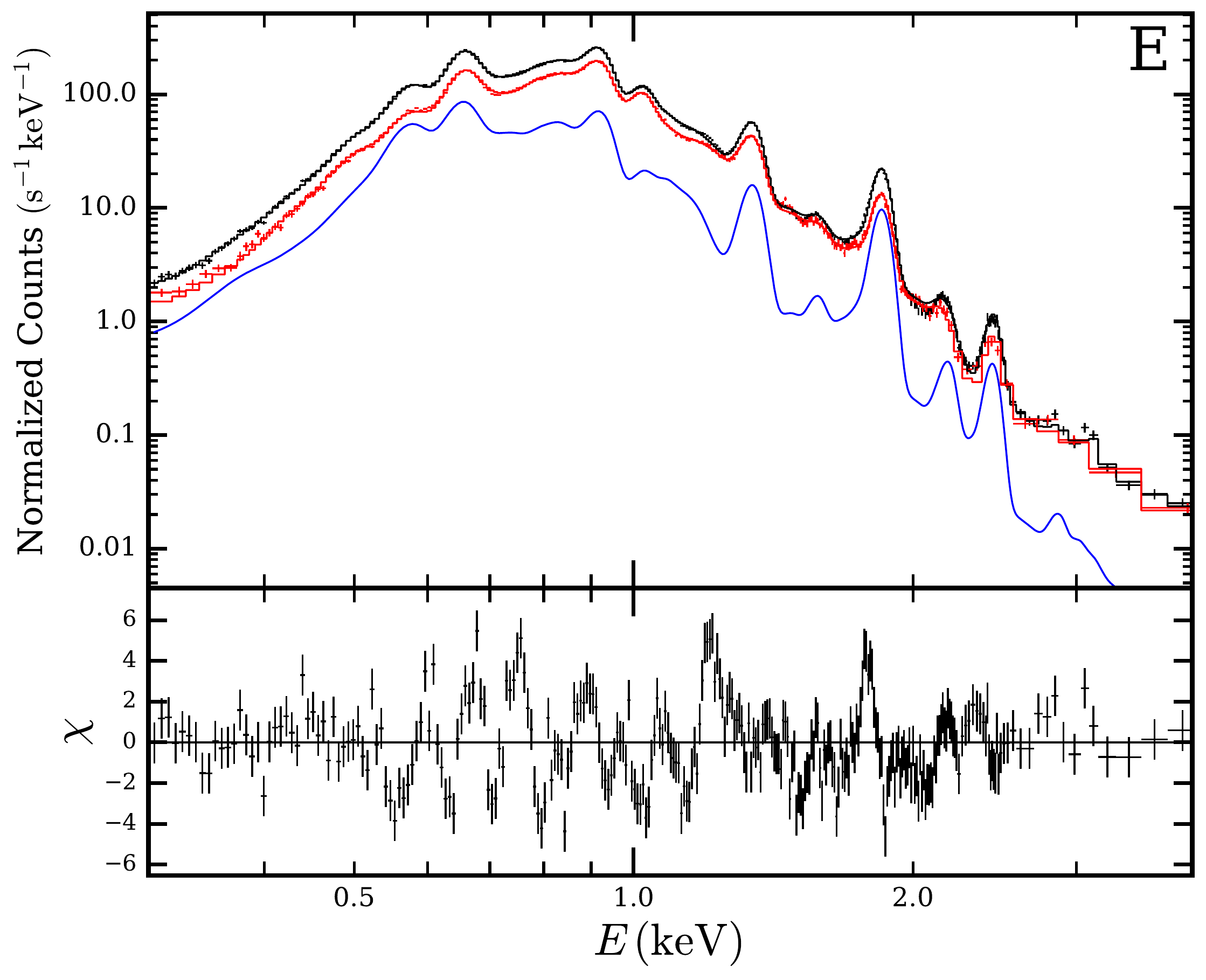} 
\includegraphics[width=6.0cm]{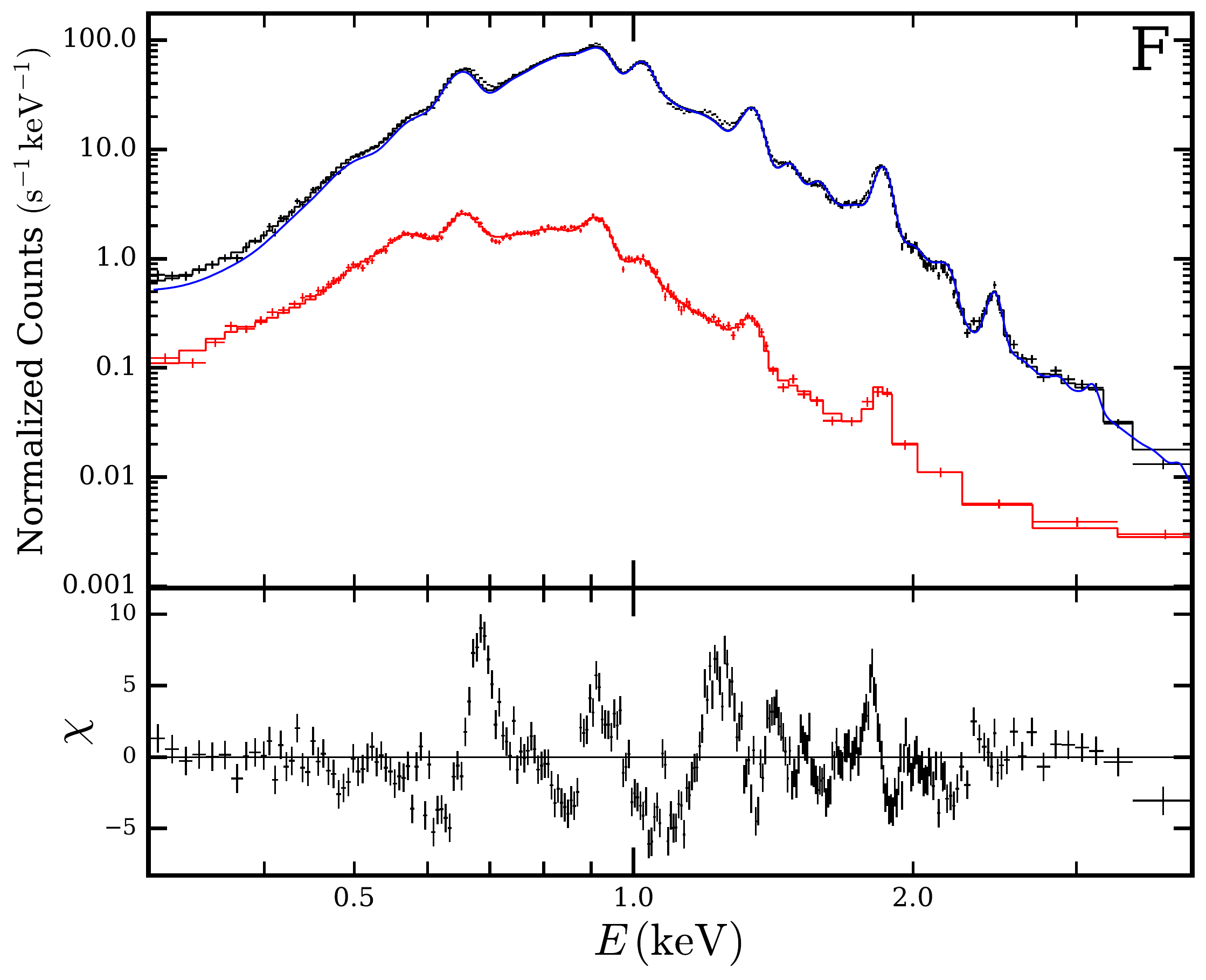} \\
\includegraphics[width=6.0cm]{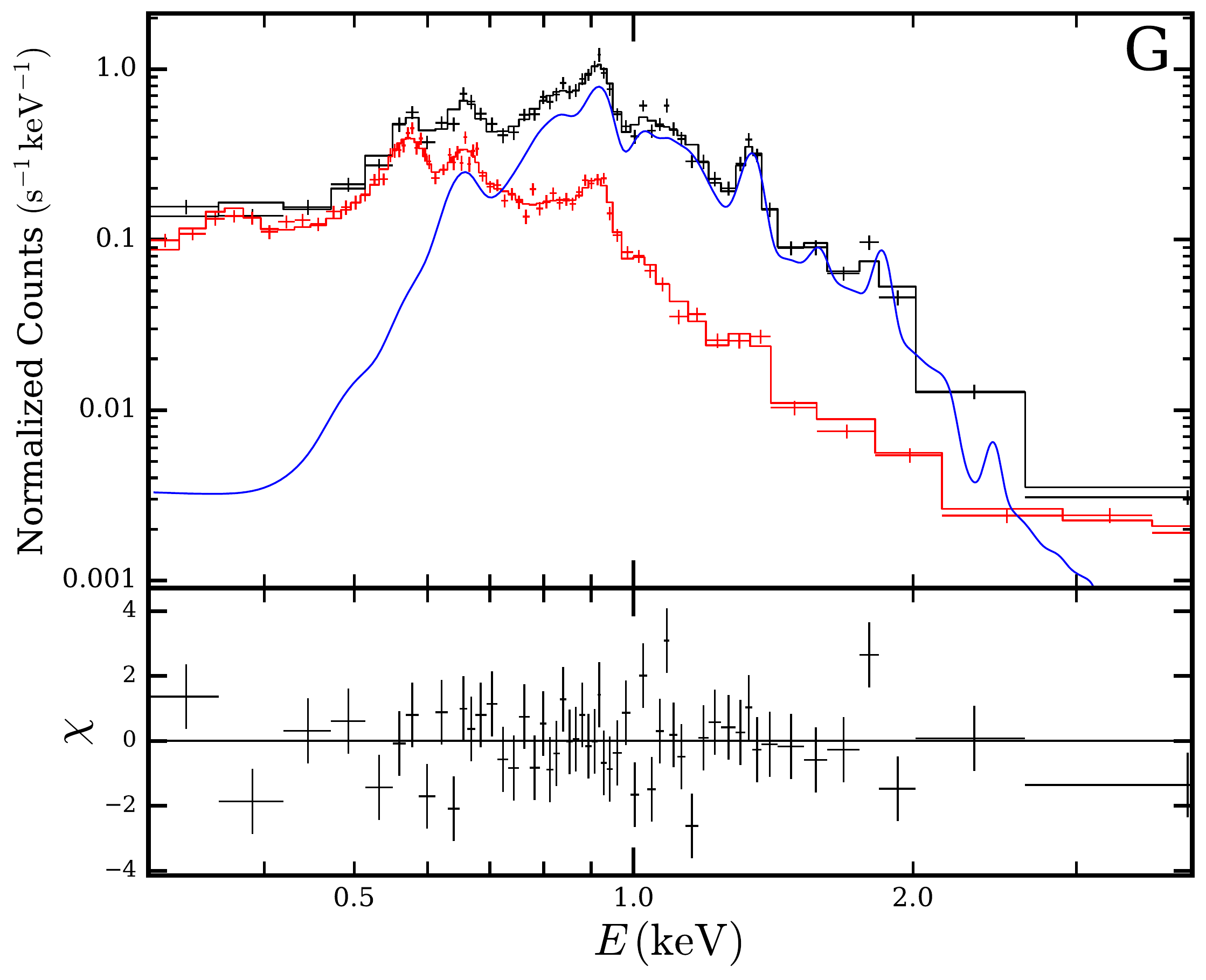}
\includegraphics[width=6.0cm]{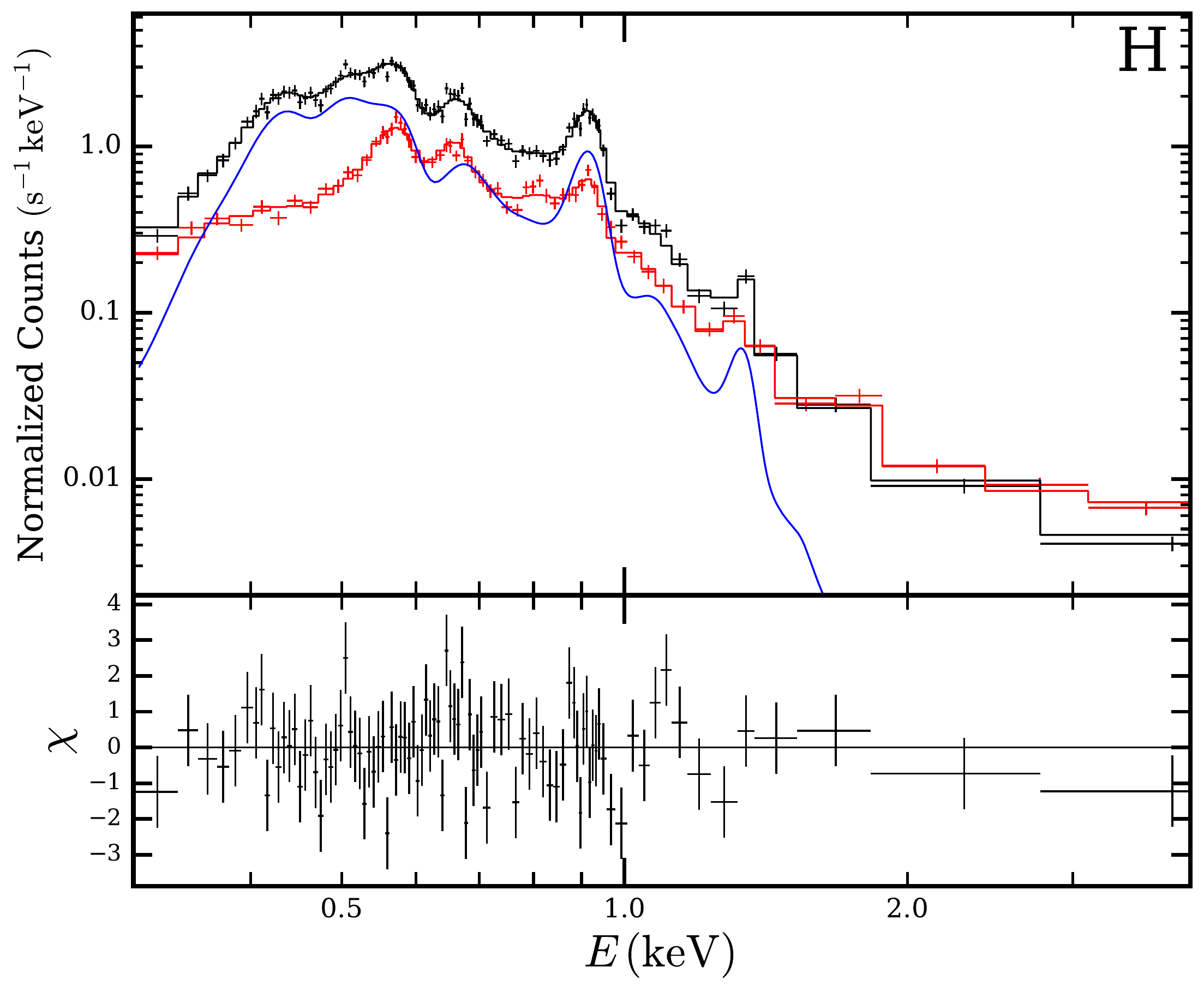} 
\includegraphics[width=6.0cm]{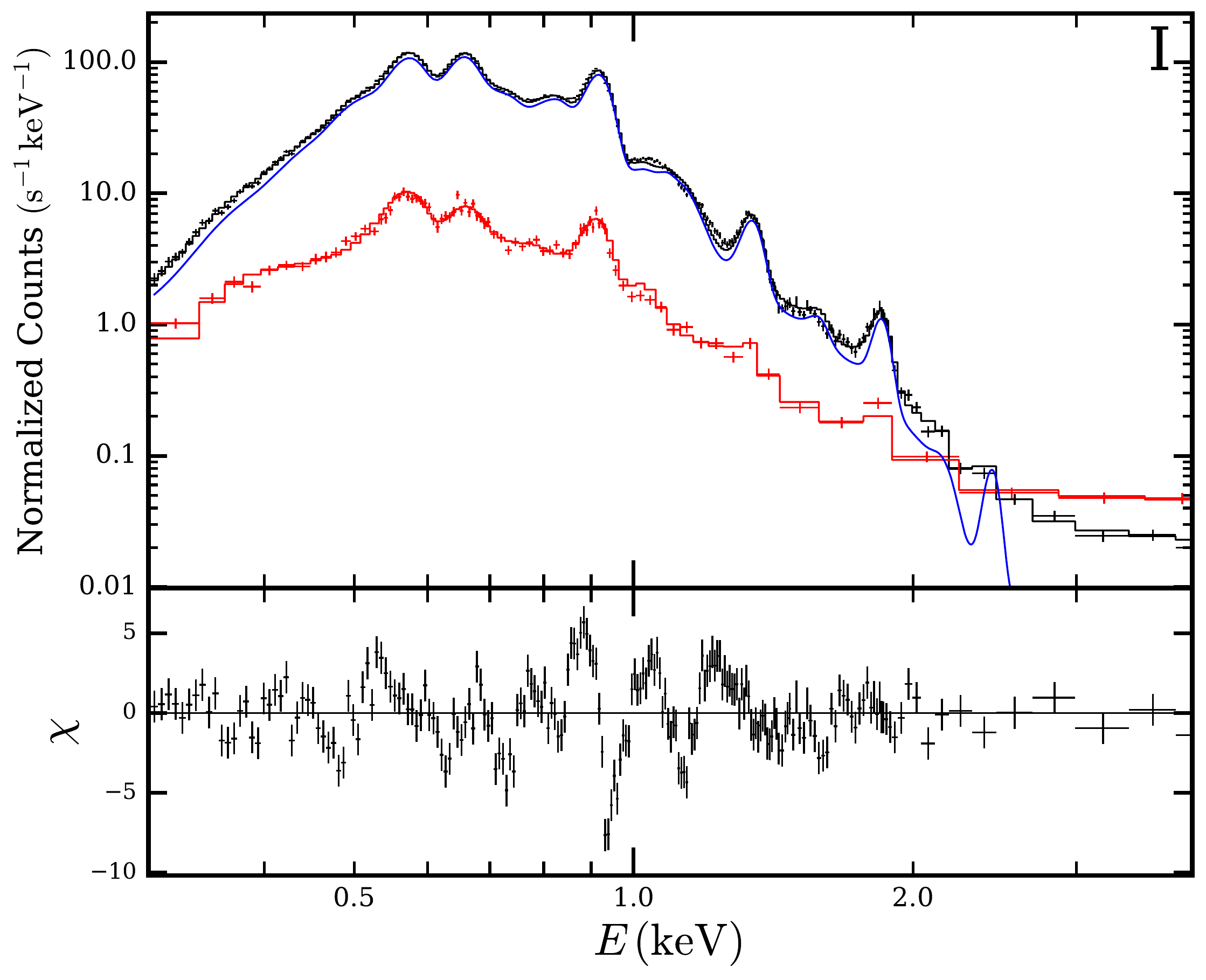} 
\caption{Detailed fits to source and background spectra extracted from the regions indicated in Fig.~\ref{DetailedFitRegions} and labelled accordingly.
In each panel, the background spectrum and binned model are displayed in red, while the total spectrum and model for the source region are shown in black. The blue line indicates the ``background-subtracted'' best-fit source model. The background spectra and models have been rescaled to represent their contribution to the source region spectrum. 
The lower part of each panel indicates the residuals of the spectrum in the source region with respect to its best-fit model for source plus background. 
The fits displayed here correspond to case (i) outlined in the text, meaning metal abundances were left free to vary within a range typical for (enriched) ISM. 
The spectra and models were rebinned, for plotting purposes only, to a minimum signal-to-noise ratio of $5$. 
}
\label{DetailedFits}
\end{figure*}

Following up on some open questions from the previous section, we now investigate in more detail the spectra of some isolated features of Puppis A, which were selected based on our images and parameter maps. 
The extraction regions used for our features are indicated in Fig.~\ref{DetailedFitRegions}. Our targets include the northeast filament (regions A and B), the compact ejecta knot (C and D), the extended ejecta-rich region (E), the hot northeast rim (F), and the cold western arc (I).
In addition, we included the southern ``hole'' of soft emission (G) to investigate the foreground absorption there, as well as an extremely faint filamentary feature at the southwest edge of Puppis A (H), which exhibits the softest visible emission associated to Puppis A.

As in the spatially resolved spectroscopic analysis described in Sect.~\ref{Spectroscopy}, the source emission was described by an absorbed plane-parallel shocked plasma, expressed as {\tt TBabs*vpshock}. To be comparable with previous studies \citep[e.g.,][]{Katsuda10,Katsuda08}, we tested two different approaches for treating elemental abundances: (i) we allowed the abundances of O, Ne, Mg, Si, S, Fe to freely vary between $0$ and $50$, reflecting ISM possibly enriched by ejecta. (ii) We fixed the oxygen abundance at a value of $2000$ \citep{Winkler85}, to allow for the possibility of pure metal ejecta, where the continuum bremsstrahlung emission is dominated by heavy elements. 

For each source region, we defined a nearby region from which we extracted a local background spectrum. As regions A$-$E are intended to single out prominent features from the surrounding ISM-dominated emission of the SNR, their background regions were chosen to approximate the local emission from Puppis A. For all other features, the regions were chosen such as to trace the ``true'' local fore- and background outside the SNR shell. 
As a background model, we used again the combination of thermal, nonthermal and instrumental components outlined in Sect.~\ref{Spectroscopy}. The spectra of the source and background regions were fitted simultaneously, with the relative normalization of the background initially tied between the two regions. After this initial fit, the X-ray background normalization was allowed to vary for the source region by up to a factor of two, to accommodate possible spatial variations in the flux of the background component. Such spatial variations may become important in particular if the background component consists of the emission of Puppis A itself.

The resulting fits to source and background spectra are displayed in Fig.~\ref{DetailedFits}. While our models are generally able to qualitatively reproduce the shape of the observed data very well, a few shortcomings become apparent: first, several of our spectra (e.g., regions E, F, I) show strong systematic deviations from their best-fit model, clearly visible because of the excellent photon statistics. This indicates that statistical errors produced with standard methods are likely underestimated with respect to realistic uncertainties of model parameters, whose dominant source is of systematic nature. 
Possible reasons for the observed fit residuals include imperfect energy or response calibration, line shifts or line broadening due to non-zero radial velocities of the plasma, the presence of multiple plasma components along the same line of sight, or shifts in the centroid of line complexes due to physical processes not described by our model. For instance, \citet{Katsuda12} found that, in the targeted regions in the east and north of Puppis A, forbidden-to-resonance line ratios in the He$\alpha$ triplets of N, O, Ne are higher than expected from purely thermal models, and require charge-exchange processes in order to be satisfactorily explained.   
A further subtle but fundamental issue of our approach is the assumption of the spectrum in the local background region being representative of the actual background contribution to the source region, which is generally reasonable, but cannot be guaranteed. 
This is especially important for those regions where source and background levels are comparable, such as A$-$E and the low- and high-energy portions of G and H, which makes them vulnerable to slight background variations. For regions A$-$E,  variations on small scales are particularly likely, as the ``background'' consists of the surrounding emission from Puppis A, which is certainly more spatially inhomogeneous than the emission outside the SNR shell. 

\begin{table*}[h!]
\renewcommand{\arraystretch}{1.5}
\caption{Best-fit parameters from modelling of the spectra shown in Fig.~\ref{DetailedFits}. \label{DetailedFitsTable}}
\centering
\resizebox{18.4cm}{!}{
\begin{tabular}{cccccccccccccc}
\hline\hline
Region&$N_{\rm H}$&$kT$&$\tau$&$\mathrm{EM}$\tablefootmark{b}&$\bar{n}_{e}$&$t_{s}^{\prime}$&$\rm O/H$&$\rm Ne/H$&$\rm Mg/H$&$\rm Si/H$&$\rm S/H$&$\rm Fe/H$&$\mathcal{C}$\tablefootmark{d} \\
&$10^{21}\,\rm{cm}^{-2} $&$\rm{keV} $&$10^{11} \rm{cm^{-3}\,s} $&$10^{12}\,\rm{cm}^{-5} $&$\rm{cm^{-3}} $&$10^3 \rm{yr} $&$ $&$ $&$ $&$ $&$ $&$ $& \\
\hline
A&$5.42_{-0.05}^{+0.12}$&$0.430_{-0.021}^{+0.010}$&$0.264_{-0.012}^{+0.027}$&$2.09_{-0.17}^{+0.31}$&$7.7 \pm 0.4$&$0.109 \pm 0.010$&$1.89_{-0.08}^{+0.09}$&$1.69_{-0.04}^{+0.10}$&$1.61_{-0.13}^{+0.11}$&$1.7_{-0.4}^{+0.4}$&...\tablefootmark{c}&$0.32_{-0.04}^{+0.04}$&1683.5 \\
A\tablefootmark{a}&$6.18_{-0.08}^{+0.10}$&$0.342_{-0.023}^{+0.004}$&$0.36_{-0.02}^{+0.06}$&$0.0038_{-0.0002}^{+0.0008}$&$0.90 \pm 0.06$&$1.25 \pm 0.16$&$2000$\tablefootmark{a}&$1520_{-50}^{+20}$&$1620_{-110}^{+90}$&$1800_{-700}^{+700}$&...\tablefootmark{c}&$140_{-40}^{+40}$&1699.0 \\
B&$5.64_{-0.29}^{+0.20}$&$0.69_{-0.07}^{+0.12}$&$0.177_{-0.028}^{+0.027}$&$0.38_{-0.11}^{+0.10}$&$6.7 \pm 0.9$&$0.084 \pm 0.018$&$3.5_{-0.3}^{+0.6}$&$3.3_{-0.4}^{+0.7}$&$3.8_{-0.4}^{+0.8}$&$3.3_{-0.6}^{+0.8}$&$<1.1$&$0.13_{-0.09}^{+0.12}$&1607.5 \\
B\tablefootmark{a}&$5.3_{-0.3}^{+0.4}$&$0.59_{-0.12}^{+0.14}$&$0.21_{-0.05}^{+0.10}$&$0.00057_{-0.00014}^{+0.00028}$&$0.76 \pm 0.14$&$0.9 \pm 0.4$&$2000$\tablefootmark{a}&$2050_{-100}^{+80}$&$2900_{-180}^{+200}$&$2900_{-500}^{+400}$&$<1100$&$<11$&1620.4 \\
C&$3.39_{-0.21}^{+0.30}$&$0.72_{-0.11}^{+0.09}$&$0.19_{-0.03}^{+0.05}$&$0.050_{-0.017}^{+0.029}$&$2.1 \pm 0.5$&$0.28 \pm 0.09$&$9_{-2}^{+5}$&$12_{-4}^{+7}$&$15_{-4}^{+6}$&$9_{-3}^{+5}$&$<16$&$<0.12$&1345.6 \\
C\tablefootmark{a}&$3.57_{-0.13}^{+0.22}$&$0.59_{-0.05}^{+0.05}$&$0.24_{-0.04}^{+0.05}$&$0.00026_{-0.00005}^{+0.00005}$&$ 0.46 \pm 0.04$&$1.6 \pm 0.3$&$2000$\tablefootmark{a}&$2790_{-130}^{+140}$&$3650_{-150}^{+220}$&$2600_{-500}^{+500}$&$4300_{-1500}^{+1600}$&$<13$&1360.7 \\
D&$4.57_{-0.10}^{+0.10}$&$1.14_{-0.09}^{+0.11}$&$0.35_{-0.03}^{+0.05}$&$0.080_{-0.023}^{+0.031}$&$2.1 \pm 0.4$&$0.52 \pm 0.11$&$16_{-4}^{+6}$&$13_{-3}^{+5}$&$14_{-4}^{+6}$&$12_{-3}^{+4}$&$8.0_{-2.0}^{+2.1}$&$<0.037$&1561.6 \\
D\tablefootmark{a}&$4.46_{-0.09}^{+0.11}$&$1.27_{-0.12}^{+0.13}$&$0.323_{-0.029}^{+0.017}$&$0.00059_{-0.00003}^{+0.00004}$&$0.509 \pm 0.016$&$2.01 \pm 0.16$&$2000$\tablefootmark{a}&$1640_{-50}^{+40}$&$1710_{-80}^{+70}$&$1410_{-60}^{+60}$&$1020_{-230}^{+240}$&$<3.0$&1593.4 \\ 
E&$3.45_{-0.07}^{+0.05}$&$0.727_{-0.008}^{+0.003}$&$0.443_{-0.020}^{+0.023}$&$1.39_{-0.12}^{+0.03}$&$0.949 \pm 0.025$&$1.48 \pm 0.08$&$3.19_{-0.11}^{+0.17}$&$4.22_{-0.27}^{+0.23}$&$4.8_{-0.3}^{+0.3}$&$17.5_{-1.1}^{+1.4}$&$20.2_{-2.3}^{+2.4}$&$2.37_{-0.15}^{+0.18}$&2687.2 \\
F&$4.103_{-0.021}^{+0.012}$&$0.7403_{-0.0015}^{+0.0007}$&$3.14_{-0.02}^{+0.06}$&$9.81_{-0.09}^{+0.01}$&$2.982 \pm 0.008$&$3.34 \pm 0.04$&$0.907_{-0.010}^{+0.016}$&$1.143_{-0.008}^{+0.011}$&$0.812_{-0.008}^{+0.009}$&$0.885_{-0.013}^{+0.014}$&$0.78_{-0.05}^{+0.05}$&$0.5551_{-0.0016}^{+0.0041}$&3460.0 \\ 
G&$11.0_{-1.0}^{+1.1}$&$0.68_{-0.08}^{+0.08}$&$0.61_{-0.13}^{+0.43}$&$0.33_{-0.07}^{+0.23}$&$0.46 \pm 0.11$&$4.2 \pm2.2$ & $0.9_{-0.4}^{+0.8}$&$0.9_{-0.2}^{+0.4}$&$0.55_{-0.10}^{+0.12}$&$0.52_{-0.11}^{+0.13}$&$0.7_{-0.6}^{+0.8}$&$0.48_{-0.11}^{+0.11}$&1811.2 \\ 
H&$4.92_{-0.15}^{+0.13}$&$0.142_{-0.018}^{+0.002}$&$0.37_{-0.14}^{+0.14}$&$20_{-5}^{+4}$&$3.0 \pm 0.3$&$0.40 \pm 0.16$&$0.072_{-0.006}^{+0.008}$&$0.39_{-0.07}^{+0.12}$&$3.7_{-1.5}^{+2.5}$&$<15$&$<13$&$<2.2$&1773.4 \\
I&$3.581_{-0.025}^{+0.027}$&$0.3511_{-0.0007}^{+0.0004}$&$0.960_{-0.015}^{+0.018}$&$21.5_{-0.5}^{+0.4}$&$1.228 \pm 0.014$&$2.48 \pm 0.05$&$0.650_{-0.015}^{+0.011}$&$0.915_{-0.021}^{+0.016}$&$0.483_{-0.011}^{+0.013}$&$1.10_{-0.05}^{+0.05}$&$8.2_{-0.7}^{+1.7}$&$0.356_{-0.009}^{+0.005}$&2552.6 \\ 
\hline
\end{tabular}
}
\tablefoot{All errors and upper limits of fitted parameters were estimated using the built-in {\tt error} command, and are given at a $1\sigma$ level. Uncertainties of derived quantities were estimated using Gaussian error propagation. 
The column $\mathcal{C}$ displays the combined Cash statistic of the best fit to source and background spectra. \\
\tablefoottext{a}{In these cases, the approach of fixing the oxygen abundance to a value of 2000 to represent pure ejecta emission (case ii), was found to provide a fit of comparable or slightly poorer quality to case (i).}
\tablefoottext{b}{This corresponds to the definition of $\rm EM$ in Xspec (see Eq.~\ref{EMEquation}), omitting the arbitrary factor $10^{-14}$. }
\tablefoottext{c}{Here, our fit was unable to provide any constraints within the allowed abundance range. }
\tablefoottext{d}{The associated number of degrees of freedom is 1550/1551 for cases (i)/(ii), respectively.}
}
\end{table*}

Nonetheless, while being aware of the above issues regarding modelling uncertainties, we believe that a discussion of the implications of our fits (see Table \ref{DetailedFitsTable} for the best-fit parameters) is warranted. 
For a few regions, approaches (i) and (ii) yielded comparable fit qualities, for all others we show only the results of the ``standard'' approach (i). 
Similarly to the previous section, we constructed density and shock age estimates for the emitting plasma from the emission measure and ionization timescale, with two modifications: first, for those features which visually appear as clumps (i.e., regions A$-$E), we assumed a line-of-sight depth of the emitting plasma similar to their extent in the image domain. This means we assumed the depth to be equal to the geometric mean of the major and minor axis of the ellipse used as extraction region.
Second, for the case of pure-metal abundances (ii), we explicitly calculated the average atomic mass per hydrogen atom and the ratio of electrons to hydrogen atoms from the measured abundances, as the usual assumptions, valid for ISM conditions, break down here. To achieve this, for each region, we calculated the ion fractions for each element from the best-fit parameters, by averaging over the flat distribution of ionization timescales from zero to its maximum, $\tau$, using the AtomDB code \citep{AtomDB}, .

The spectra of regions A and B, extracted at the northeast filament, clearly confirm the low ionization age of the plasma that we found in Sect.~\ref{Spectroscopy}. Comparing our results for case (i) with the maps in Fig.~\ref{TesselationImage}, our isolated treatment of this region and the more realistic assumption of a compact three-dimensional structure yield larger density estimates on the order of $10\,\si{cm^{-3}}$, which imply smaller shock ages $\sim 100\,\si{yr}$. Such recent  interaction with a shock seems to agree well with the -- to our knowledge unpublished -- suspected flux decline over time in this filament \citep[see][]{Katsuda_Prop}, which would be a natural expectation for a comparatively ``young'' feature evolving on a short timescale. 

Moreover, it is interesting to note that both regions A and B tend toward a quite large absorption column around $N_{\rm H} \approx 5.5 \times10^{21}\,\si{cm^{-2}}$, much larger than in adjacent regions, where $N_{\rm H} \approx 3\times10^{21}\,\si{cm^{-2}}$ (weakly visible also in Fig.~\ref{TesselationImage}). 
This is also found if one keeps the relative background normalization fixed, and is therefore not an artifact of background over-subtraction at low energies. Furthermore, \citet{Katsuda10} noted a similar behavior in their analysis before fixing $N_{\rm H}$ to a smaller, more ``reasonable''  value, which is why we consider it a possibility that the enhanced absorption may indeed be physical (see Sect.~\ref{AbsDiscuss}).
Our higher absorption has a considerable effect on the measured abundances: we generally recover supersolar abundances of O, Ne, Mg, Si in both regions, with the northern end of the filament (region B) exhibiting somewhat higher values. However, the ratios $\rm Ne/O$ and $\rm Mg/O$ are considerably below those found by \citet{Katsuda10}, who obtained values approximately twice solar. This is likely due to the higher absorption suppressing the soft oxygen line emission in the model spectrum, requiring an increased abundance to match the observed line flux.

Regions C and D represent the southern and northern part of the compact ejecta knot. In both cases, a fit with pure-ejecta abundances (ii) provides only a slightly worse fit compared to ISM abundances (i). While it is hard to make exact statements about temperature and ionization age due to the difficult access to associated continuum emission, the northern part appears to exhibit a hotter plasma, somewhat closer to CIE. 
Furthermore, we find that in the northern part of the clump, oxygen seems to be the most strongly enhanced element, while in the south, neon and magnesium (and possibly silicon) are more concentrated.
Interestingly, we find no need to include emission from iron in any of the assumed scenarios in either region. 
The spectrum of region E, corresponding to the more extended ejecta-rich region, strongly contrasts those of regions C and D, as it clearly exhibits very strong silicon line emission, consistent with the extreme measured abundance thereof. At the same time, the abundances of O, Ne, Mg, while also found to be enhanced, are around a factor of $3-5$ lower. This provides a clear confirmation of the spatial separation of silicon from lighter elements in the ejecta of Puppis A. 

Spectrum F corresponds to the northeast rim, which stands out clearly as the hardest extended source of emission within Puppis A.  
Our fits confirm the high plasma temperature around $0.75\,\si{keV}$ and large ionization age here, which in combination lead to a pronounced tail toward high energies in the spectrum \citep[see also][]{Krivonos21}. 
In this context, it is interesting to note that this region of hard emission appears to coincide approximately with the peak of emission of Puppis A in GeV gamma rays \citep{Xin17}. However, our observation does not appear to suggest any indication of non-thermal emission, which would be expected if there was a contribution to the X-ray emission by synchrotron radiation from accelerated particles. 
In addition to providing a quantitative temperature measurement, we recover weakly enhanced elemental abundances in our region (compared to the median over the SNR) as well as a ratio of $\rm Ne/O \approx 1.25$, which is somewhat lower than in other ISM-dominated parts of the remnant (see Fig.~\ref{MetalRatioImage}).

As expected, the spectrum of Region G, the ``southern hole'' in soft emission, is strongly absorbed, with source emission only being detected above around $0.6\,\si{keV}$. While most physical parameters are not remarkable, the obtained column density of $N_{\rm H} \approx 11\times10^{21}\,\si{cm^{-2}}$ is far higher than in any other region of Puppis A, consistent with the peak in foreground dust emission \citep{Arendt10,Dubner13}.
In addition, one should note that the measured temperature of around $0.7\,\si{keV}$ is quite large compared to the immediate surroundings as apparent in Fig.~\ref{TesselationImage}, which may point towards a possibly problematic fit. For instance, a possible superposition of different absorption column densities in the same spectrum would tend to mimic an increased plasma temperature. The reason for this is that the low-energy portion of the spectrum, which would be dominated by the least absorbed component, would force the fit towards a small value of $N_{\rm H}$, due to the exponential effect of absorption on the observed spectrum. The additional hard emission introduced by more heavily absorbed components would then lead to an artificial increase in the fitted plasma temperature for a single-component model \citep[for an example of this effect, see][]{Locatelli22}.
In addition, the scenario of different superimposed absorption columns is also consistent with the expectation for a compact absorbing cloud \citep{Arendt10}, which would likely not have uniform optical thickness across region G.
In conclusion, the true maximum column density towards the ``southern hole'' may be significantly larger than $11\times10^{21}\,\si{cm^{-2}}$.

Region H exhibits a quite curious spectrum, whose best-fit parameters appear implausible, as they tend toward a strongly absorbed and comparatively cold plasma with strong departure from CIE. Nevertheless, the observed spectral shape is distinctly different from any other spectrum in Fig.~\ref{DetailedFits}, as the source emission is the brightest around $0.4-0.6\,\si{keV}$ but exhibits an extremely sharp cutoff toward lower energies. This appears to make the high column density and low temperature at least somewhat plausible. 
However, also the fitted elemental abundances appear quite peculiar: while the large magnesium abundance could likely be resolved by including a second, hotter plasma component in emission, the extremely low measured oxygen abundance is remarkable, especially since it is robustly recovered by the fit, independent of the exact modelling approach for source or background. 

Finally, region I was defined in order to test for the presence of enhanced sulfur abundances in the western arc, as these were found to often diverge in previous fits with low signal-to-noise.  
Our best-fit single-component model recovers the previously found low plasma temperature, confirming that the region around the western arc, on average, hosts significantly colder plasma than the vast majority of Puppis A. It is interesting to note that the western arc appears distinct from the rest of the SNR, not only from a spectral but also from a morphological point of view, given its almost circular structure somewhat separate from the rest of the shell (see Fig.~\ref{PuppisImage}). A possible explanation for the formation of such ``ear''-like structures in SNRs could be the expansion of the shock wave into a non-isotropic ISM, which may for instance be formed by an equatorially concentrated wind from the progenitor \citep{Chiotellis21}.  

Apart from the low plasma temperature, our spectral fit of region I indeed yields a strongly enhanced sulfur abundance at around eight times solar. 
This result should be treated with extreme caution, however. For instance, if we allow for a second plasma component in emission (with abundances tied to the first), we obtain an alternative fit with an improved statistic by around $\Delta\mathcal{C} \approx  -200$. This second thermal component is hotter than the primary one at a temperature of $0.5\,\si{keV}$, and completely removes the need for any sulfur line emission in the model. 
The conclusion is therefore that the putative sulfur enhancement, while formally statistically significant in our basic model, may in reality be an artifact of a hard tail of the spectrum which is incompatible with the lower temperature of a single emission component.  
This case illustrates that, in theory, recreating realistic physical conditions of shocked plasma in SNRs would often require a complex multi-component treatment of the emission, or even a non-trivial continuous distribution of temperatures and ionization timescales. However, when in practice fitting present-day CCD-resolution spectra, it is often challenging to convincingly disentangle even two emission components without imposing significant restrictions on the allowed parameter space.

\subsection{The spectrum of the CCO \label{CCOSpec}}

\begin{table*} 
\renewcommand{\arraystretch}{1.5}
\caption{Best-fit parameters for single (BB) and double (2BB) blackbody fits to the spectrum of the CCO. \label{CCOFitResults}}
\centering
\begin{tabular}{ccccc}
\hline\hline
 & \multicolumn{2}{c}{This work} & \multicolumn{2}{c}{\citet{HuiBecker06}} \\
 & BB & 2BB & BB & 2BB \\
\hline
$N_{\rm H}\, (10^{21}\,\si{cm^{-2}})$ & $3.14^{+0.14}_{-0.14}$ & $4.65^{+0.83}_{-0.53}$ & $2.67^{+0.09}_{-0.09}$ & $4.54^{+0.49}_{-0.43}$ \\
$kT_1 \, (\si{keV})$ & $0.347^{+0.005}_{-0.005}$ & $0.404_{-0.026}^{+0.039}$ &$0.372^{+0.003}_{-0.003}$ & $0.434^{+0.024}_{-0.017}$ \\
$R_1 \, (\si{km})$\tablefootmark{a} & $0.89_{-0.03}^{+0.03}$ & $0.56_{-0.17}^{+0.16}$ & $0.73^{+0.02}_{-0.01}$ & $0.44^{+0.07}_{-0.09}$\\
$kT_2 \, (\si{keV})$ & ... & $0.215^{+0.040}_{-0.040}$ & ... & $0.225^{+0.026}_{-0.022}$ \\
$R_2 \, (\si{km})$\tablefootmark{a} & ... & $1.95_{-0.53}^{+0.99}$ & ... & $1.94^{+0.66}_{-0.44}$\\
$F_{a} \, (10^{-12}\,\si{erg.s^{-1}.cm^{-2}})$\tablefootmark{b} & $4.54^{+0.04}_{-0.07}$ & $4.70^{+0.07}_{-0.07}$ & $4.42$ & $4.88$\\
$F_{u} \, (10^{-12}\,\si{erg.s^{-1}.cm^{-2}})$\tablefootmark{c} & $6.66^{+0.12}_{-0.12}$ & $8.78_{-0.81}^{+1.47}$ & $6.05$ & $8.94$ \\

\hline
\end{tabular}
\tablefoot{Errors were estimated using the built-in Xspec command. \\
\tablefoottext{a}{Size of the circular emitting region converted assuming a distance of $1.3\,\si{kpc}$.}
\tablefoottext{b}{Observed (i.e., absorbed) flux of the source in the $0.5-10\,\si{keV}$ range.}
\tablefoottext{c}{Absorption-corrected flux of the source in the $0.5-10\,\si{keV}$ range.}
}
\end{table*}

For completeness, we dedicate some further attention to the CCO of Puppis A, RX J0822$-$4300, visible as a hard point source in imaging. 
The timing of the CCO has been investigated in detail in the past \citep[e.g.,][]{Gotthelf13}, 
and the temporal resolution of $50\, \si{ms}$ delivered by eROSITA is insufficient to provide any new insights in this respect. 
However, we can use its spectrum to verify if the results of our measurements are generally sensible, as variations of the spectrum are not expected on timescales of years to even decades. 
We thus extracted its spectrum from a circular aperture of a radius of $30\arcsec$ centered on the CCO, while we determined the local background from a concentric annulus with inner and outer radii of $50\arcsec$ and $90\arcsec$. 
For modelling of the background, we used the same approach as in the previous section. 
Following \citet{HuiBecker06}, we used absorbed single (BB) and double (2BB) blackbody models to describe the emission of the CCO, from which we obtained the fit parameters displayed in Table \ref{CCOFitResults}.

For the BB model, our measurement appears statistically discordant with the study of \citet{HuiBecker06}, who found a slightly hotter and less absorbed blackbody with somewhat smaller emitting area than our analysis suggests. 
However, it is important to keep in mind the systematic effect of the worse spatial resolution of eROSITA compared to {\it Chandra} and {\it XMM-Newton}, as the stronger blending of source and background likely outweighs the statistical errors derived in Xspec. Furthermore, since the BB model is likely not the optimal description of the spectrum \citep{HuiBecker06}, the smaller hard response of eROSITA leads to a weaker weighting of the high-energy tail of the spectrum, which naturally entails a lower measured effective temperature.

In this light, it is quite reassuring to notice that all measured parameters for the likely more accurate 2BB model agree within their statistical uncertainties, as we recover two blackbody components of temperatures around $0.2$ and $0.4\,\si{keV}$ with emission areas significantly smaller than the neutron star surface. Furthermore, all four fits described in Table \ref{CCOFitResults} yield observed broad-band fluxes of the CCO that show a convincing agreement to within ten percent. A similar agreement can be found for the intrinsic fluxes when comparing identical models fitted here and in \citet{HuiBecker06}. The apparent disagreement between unabsorbed fluxes of the BB and 2BB models can be attributed to the vastly different measured column densities.
Apart from confirming the results of \citet{HuiBecker06}, we believe that this brief analysis demonstrates that the absolute calibration of the eROSITA response in our observation is reliable up to a level of approximately ten percent.

\section{Discussion \label{Discussion}}

\subsection{X-ray absorption toward Puppis A \label{AbsDiscuss}}
While often regarded as somewhat of a nuisance parameter when analyzing spectra, the X-ray absorption toward Puppis A warrants some further discussion for several reasons:
first, we found that the absorption column density shows strong variations of at least a factor five over the SNR's extent, from around $2\times10^{21}\,\si{cm^{-2}}$ in the east and northwest to around $11\times10^{21}\,\si{cm^{-2}}$ in the very south of Puppis A (Sect.~\ref{DetailedFitSection}). Therefore, a spatially integrated treatment of the emission from this SNR is not adequate, as it would not only combine emission from regions of different temperatures but also of different absorption columns. Strictly speaking, the resulting integrated spectrum is therefore not describable with a single-component model, and the derived parameters, such as temperature and intrinsic flux, can thus be strongly biased (see Sect.~\ref{Spectroscopy}).

Second, with our analysis, we have shed some further light on the question of what causes the peculiar strip of hard emission crossing Puppis A from northeast to southwest (see Fig.~\ref{PuppisImage}). 
Numerous previous studies \citep{Aschenbach93,HuiBecker06, Dubner13} have, with good reason, argued that this strip is exclusively caused by an absorbing filament, visible in \ion{H}{i} emission. 
However, our analysis suggests that the observed hard strip is not only due to foreground absorption, but caused by a combination of strong absorption in the southwest and elevated plasma temperatures in the center and northeast of the SNR. Apart from our spectral modelling, this is further supported by the fact that the northeast rim of Puppis A becomes the increasingly dominant source of emission above $2\,\si{keV}$ \citep[see bottom row of Fig.~\ref{NarrowBandImages} and][]{Krivonos21}, where one would expect the spectral hardness to be mostly dependent on temperature rather than absorption. 

Finally, both our work and the study of \citet{Katsuda10} found locally enhanced absorption toward the northeast filament of $N_{\rm H} \sim 5.5\times10^{21}\,\si{cm^{-2}}$. \citet{Katsuda10} chose to discard this finding, since enhanced absorption appears counterintuitive for a feature that exhibits softer emission than its surroundings. However, since an elevated hydrogen column density was independently detected with two different instruments, we believe that it is at least a possibility that filaments such as this one may exhibit localized enhanced X-ray absorption. 
Additional support for this idea comes from the Fig.~1 in \citet{Arendt10}, an overlay of mid-infrared (MIR) and X-ray emission of Puppis A, which shows superb agreement between the two regimes for a large fraction of the SNR, including the northeast filament. Thus, as the MIR is believed to trace the emission of shocked or heated dust, we suggest that the material that contributes additional small-scale absorption may actually be located within Puppis A, and could likely be (possibly destroyed) dust within the northeast filament. 

\subsection{The distribution of ejecta} \label{DiscEjDist} 
The only previous spatially resolved abundance measurements covering the entirety of Puppis A were published by \citet{Luna16}. 
However, the map of oxygen abundance displayed in their work does not show any clear peak close to the ejecta knot visible in our corresponding plot in Fig.~\ref{TesselationImage}, likely due to the much larger size of their spatial bins.
Furthermore, the average abundances obtained in their study are significantly lower than in ours. As both studies used the same reference abundance table, the reason for this is unclear, but may be related to different model versions or spectral analysis packages used. 

Much better agreement exists between our results and the study of the eastern half of Puppis A by \citet{Hwang08}, concerning both the peak location and the average abundance values. 
First, we confirm their observation of overall subsolar abundances, finding median abundances ranging between around 0.4 for iron, and around 0.8 for silicon. Furthermore, we do not detect any significant large-scale oxygen enrichment to supersolar values, contrary to the natural expectation for an SNR that appears oxygen-rich in the optical \citep{Winkler85}. 
The generally low iron abundance and the lack of well localized iron enhancements throughout Puppis A could indicate a lack of iron ejecta enriching the ISM. Alternatively, the majority of iron ejecta may be in physical conditions outside the X-ray-emitting regime, as, for instance, they may not have been overrun by the reverse shock yet. 

A somewhat special case is neon, which shows the highest median abundance of all metals in ISM-dominated regions. In the light of bright dust emission in Puppis A, a promising explanation for this, introduced by \citet{Hwang08}, is the depletion of the other elements onto dust grains. Such dust depletion would reduce their measured X-ray abundances while preserving those of neon at around solar. 
The fact that the oxygen-to-neon ratio seems to be enhanced in the hot plasma at the northeast rim might be relevant in this context, as one could imagine that this may be a signature of a lower fraction of dust depletion, possibly due to its prior destruction. Of course, alternative explanations, such as a large-scale enrichment of the region with oxygen-rich ejecta material or intrinsic abundance gradients in the local ISM, are also conceivable.

A further observation by \citet{Hwang08}, uncertain at the time due to the limited spatial resolution and statistics of the observation, was that of apparently different spatial distributions of silicon and oxygen within the ejecta enhancement discovered by them. Thanks to the improvement in resolution and statistics provided by our data set, we are not only able to confirm this hypothesis, but can also see that the lighter elements (oxygen, neon, and magnesium) are considerably more concentrated than the heavier elements (silicon and sulfur), which show an elongated distribution oriented along the northeast-southwest direction. 
This is the first clear confirmation of non-uniform mixing of hydrostatically and explosively synthesized ejecta elements on large scales in Puppis A. This is especially interesting since the heavier elements, synthesized toward the interior of the star, appear to be preferentially located further out in the SNR. 

A less drastic case of non-uniform element distribution is detected within the compact ejecta knot that was previously investigated by \citet{Katsuda08,Katsuda13}. 
Our results of spectral fits to the northern and southern portion of the clump (Sect.~\ref{DetailedFitSection}) are in broad qualitative agreement with those previous results, given the different approaches and models used. We find that the two parts show different chemical compositions, with the measured abundance ratios of neon and magnesium to oxygen being higher in the south than in the north. This observation is somewhat counterintuitive, as the southern part is considered to be associated with the oxygen-rich optically emitting ``$\Omega$ filament'' \citep{Katsuda08,Winkler85}. This may indicate that a larger fraction of oxygen in the southern portion of the clump has cooled to temperatures below the X-ray emitting regime.
A further characteristic separating the two portions of the ejecta knot is the fact that the southern part shows much stronger MIR emission than the north \citep{Arendt10}. This indicates a higher concentration of heated dust in the south, which may be cospatial with the ejecta observed in X-rays or the optical.  
Finally, we note that our measurements and those of \citet{Katsuda13} show little evidence for significant iron line emission in either part of the knot, indicating that its major constituents are in fact lighter elements, and any explosively synthesized iron, if present, is likely subdominant.  

Our elemental abundance maps show only a single region with strong enhancements of O, Ne, Mg, Si, S, with little indication of other X-ray emitting ejecta clumps throughout the remainder of the SNR. Therefore, a natural question to ask is why only such a small amount of ejecta is visible in Puppis A. 
In the context of this question, it is interesting to note that all detected compact ejecta clumps appear to be strongly underionized, implying relatively recent shock interaction. This is a natural expectation if ejecta material shocked earlier in time has already become unidentifiable as such. 
Apart from the advanced age of Puppis A, which implies that any X-ray emission from ejecta is superimposed with a dominant ISM-driven component, possible explanations for the apparent lack of ejecta include their cooling or the destruction of compact ejecta clumps.
While the cooling of ejecta out of the X-ray-emitting regime is expected to operate slowly, the destruction of ejecta clumps after their interaction with the reverse shock may occur quite rapidly. We can roughly estimate the ``clump''-crushing timescale for Puppis A, in analogy to the characteristic timescale for the interaction of shock waves with interstellar clouds \citep{Klein94}: assuming an initial density contrast between ejecta clump and ambient unshocked medium $\chi \lesssim 10$,  
a characteristic clump radius $a_0\sim0.2\,\si{pc}$, similar to the apparent size of the observed ejecta knots, and a reverse shock velocity on the order of $v_{s}\sim1000\,\si{km.s^{-1}}$ \citep{Katsuda13}, the clump-crushing timescale $t_{\rm cc}=\chi^{1/2}\,a_0/v_{s}$ evaluates to $t_{\rm cc} \lesssim 600\,\si{yr}$.
Since a clump is typically destroyed on the order of a few $t_{\rm cc}$ \citep{Klein94}, ejecta knots overrun by the reverse shock early-on in the evolution of Puppis A are expected to have significantly fragmented. The implied reduction of the density contrast between clump and ambient medium likely hampers the identification of such older ejecta-rich features in X-rays.

If we assume our abundance maps to be approximately representative of the true ejecta distribution, they allow for an interesting comparison with the kinematics of the explosion: 
the CCO RX J0822$-$4300 has a quite precisely measured proper motion \citep{Mayer20}, which, assuming a distance of $1.3\,\si{kpc}$, implies a transverse velocity of around $500 \,\si{km.s^{-1}} $ toward southwest (position angle $\phi_0 \approx 248^{\circ}$ east of north). In a hydrodynamic kick scenario \citep[e.g.,][]{Wongwa13}, one expects that heavy-element ejecta would have been expelled preferentially in the opposite direction of the recoil experienced by the neutron star, which our abundance maps seem to qualitatively support. In particular, the silicon map in Fig.~\ref{TesselationImage} indicates that close to all X-ray emitting silicon-rich ejecta are located in the northeast quadrant. 

A more quantitative comparison was performed by \citet{Katsuda18}, who performed a linear decomposition of the X-ray emission of Puppis A into physical components, aiming to isolate regions rich in intermediate-mass-element (IME) ejecta such as silicon. They found a general agreement between the direction of neutron star motion and the distribution of these elements.
However, upon closer inspection, it appears as if their method picks up part of the region of hot plasma at the northeast rim as IME-enriched, which is not confirmed by our analysis. Instead, this may be an artifact of enhanced silicon line emission introduced by the comparably high plasma temperature there, rather than enhanced abundances.

If we perform a computation of the center of mass of the silicon abundance map in Fig.~\ref{TesselationImage} (considering only clearly silicon-rich bins with $\rm Si/H > 1.2$ and weighting each region according to the measured abundance), we obtain the location $(\alpha, \delta) = (08^{\rm h}22^{\rm m}38^{\rm s}, -42^{\circ}47^{\prime}40^{\prime\prime})$. This estimate of the centroid of IME ejecta lies at a position angle of $11^{\circ}$ east of north from the commonly adopted center of expansion \citep{Winkler88}. The deviation from the ideally expected recoil direction is around $57^{\circ}$, which is a much stronger discrepancy than apparent in \citet{Katsuda18}. 
We note however that this computation assumes a perfectly well known SNR center, which in reality is highly dependent on input assumptions \citep[see][]{Mayer20}. Furthermore, the observed distribution of silicon-rich ejecta may well be biased by the location of the reverse shock, as for a mature SNR such as Puppis A, only recently heated ejecta are expected to be visible.   
Nonetheless, an observation of IME-rich ejecta in a direction apparently unrelated to the expectation from a neutron star recoil scenario is not unheard of.
For instance, the SNR N49 in the Large Magellanic Cloud was found to exhibit a deviation around $70^{\circ}$ between the computed centroid of IME ejecta and the expectation from the apparent recoil direction of its magnetar \citep{Katsuda18}.

Optical observations exhibit a qualitatively similar picture concerning the prevalence of ejecta, as oxygen-rich knots are found only in the northeast quadrant of Puppis A. Furthermore, they seem to possess an overall momentum directed in a direction roughly opposite that of the CCO \citep{Winkler88,Winkler07}.  
An important remark in this context is that, ideally, one would investigate the distribution of heavy ejecta elements, such as iron or nickel, as those are expected to show the greatest anisotropies introduced by strong kicks \citep{Wongwa13}. However, as this and previous works have shown, these elements are notoriously difficult to trace for Puppis A.  

A final interesting, though at this point uncertain, piece of evidence is the shrapnel-like feature displayed in Fig.~\ref{Bullet}, whose position with respect to the neutron star is almost opposite its proper motion direction. Its location outside the blast wave as well as the shape of the associated shock front visible in Fig.~\ref{Bullet} seem to suggest that this feature is caused by a dense clump propagating into the ISM, and experiencing reduced deceleration due to its comparatively high density. 
A possible interpretation in analogy to the Vela shrapnels \citep[e.g.,][]{Miyata01} would be that this clump is composed of fast and dense ejecta material, which, given its direction of motion, would fit well into a hydrodynamic recoil scenario. 
However, comparable features observed in the SNR RCW 103, which is similarly old as Puppis A, were found to show unremarkable abundances, contradicting an interpretation as ejecta clumps \citep{Frank15,Braun19}.

\subsection{Estimating the mass of ejecta clumps and swept-up ISM \label{MassEstimates}}

\begin{table*}
\renewcommand{\arraystretch}{1.5}
\caption{Estimated masses of features containing ejecta from spectral fits. \label{MassTable}}
\centering
\begin{tabular}{ccccccc}
\hline\hline
Region&$M_{\rm tot}$&$M_{\rm O}$&$M_{\rm Ne}$&$M_{\rm Mg}$&$M_{\rm Si}$&$M_{\rm Fe}$ \\
&$10^{-3} M_{\odot} $&$10^{-3} M_{\odot} $&$10^{-3} M_{\odot} $&$10^{-3} M_{\odot} $&$10^{-3} M_{\odot} $&$10^{-3} M_{\odot} $\\
\hline
A&$66 \pm 4$&$0.69 \pm 0.05$&$0.118 \pm 0.008$&$0.064 \pm 0.006$&$0.075 \pm 0.020$&$0.025 \pm 0.004$ \\
A\tablefootmark{a}&$ 16.2 \pm 1.0$&$ 11.2 \pm 0.7$&$ 1.62 \pm 0.11$&$ 0.99 \pm 0.09$&$ 1.3 \pm 0.5$&$ 0.17 \pm 0.05$\\
B&$13.8 \pm 1.9$&$0.26 \pm 0.05$&$0.048 \pm 0.010$&$0.031 \pm 0.007$&$0.031 \pm 0.008$&$< 0.0037$ \\
B\tablefootmark{a}&$ 3.2 \pm 0.6$&$ 2.0 \pm 0.4$&$ 0.39 \pm 0.07$&$ 0.32 \pm 0.06$&$ 0.36 \pm 0.09$&$ <0.0011$  \\
C&$6.0 \pm 1.4$&$0.27 \pm 0.12$&$0.07 \pm 0.03$&$0.050 \pm 0.021$&$0.036 \pm 0.019$&$<0.0005$ \\
C\tablefootmark{a}&$ 2.61 \pm 0.25$&$ 1.52 \pm 0.15$&$ 0.40 \pm 0.04$&$ 0.31 \pm 0.03$&$ 0.24 \pm 0.06$&$ <0.0010$ \\
D&$10.1 \pm 1.7$&$0.82 \pm 0.29$&$0.13 \pm 0.05$&$0.075 \pm 0.028$&$0.072 \pm 0.026$&$<0.00031$ \\
D\tablefootmark{a}&$ 4.38 \pm 0.14$&$ 3.07 \pm 0.10$&$ 0.479 \pm 0.020$&$ 0.288 \pm 0.015$&$ 0.269 \pm 0.014$&$ < 0.0005$ \\
E&$362 \pm 10$&$6.2 \pm 0.3$&$1.57 \pm 0.10$&$1.03 \pm 0.08$&$4.3 \pm 0.3$&$0.97 \pm 0.07$ \\
\hline
\end{tabular}

\tablefoot{The total and elemental masses shown here were derived from the best-fit parameters shown in Table \ref{DetailedFitsTable}. \\
\tablefoottext{a}{Here, the best-fit parameters with the oxygen abundance fixed to a value of 2000 were used.}
}
\end{table*}

In Sect.~\ref{Spectroscopy}, we have obtained the emission measure distribution across the whole SNR. This allows us to perform a rough estimate of the swept-up mass contributing to thermal X-ray emission in the forward shock, by integrating the product of average density and region volume over all analyzed regions associated to Puppis A.  
In order to improve the quantitative estimate of the electron density in the individual regions, it is necessary to make a more realistic assumption for the line-of-sight distribution of emitting plasma than the assumed uniform distribution over $20\,\si{pc}$ that entered the map in  Fig.~\ref{TesselationImage}.

We made the simplifying assumption that the shell of Puppis A is spherically symmetric, with a radius corresponding to the approximate observed diameter of $56\arcmin$ at a distance of $1.3\,\si{kpc}$, implying a shockwave radius of $r \approx 10.6\,\si{pc}$. Further, we assumed that the density along the radial direction follows a profile typical for an SNR in the Sedov-Taylor phase \citep{CoxAnderson82}. Using this approach, we found numerically that the quantity that effectively enters the mass and density computations (see Eq.~\ref{ne}), the product of filling factor and line-of-sight depth $D_{\rm LoS}f$, is expected to range between around $5$ and $9\,\si{pc}$ over the majority of the SNR.
In conjunction with the assumption of a mean atomic mass per hydrogen atom of $1.4\,m_{\rm p}$, typical for cosmic abundances, we obtain realistic estimates of the density in each region, and from that an estimate for the total emitting plasma mass of 
\begin{equation}
    M_{\rm ISM} = (78-87)\,\left ( \frac{d}{1.3\,\si{kpc}} \right )^{5/2}\,M_{\odot}.
\end{equation}
The given uncertainty range was obtained by comparing the results of assuming different effective SNR extents and of using different Voronoi bin sizes. Formally derived statistical error bars would be much smaller than this heuristic systematic uncertainty range due to the large number of contributing regions.

On an elementary level, our result confirms that the overall mass budget of Puppis A is indeed dominated by swept-up ISM rather than by ejecta, which is a key assumption of the Sedov-Taylor model. Of course, this is unsurprising, as the blast wave is likely to have been decelerated substantially by heavy interaction with ISM at multiple locations. 
Using the approximate present-day radius of the shock wave $r \approx 10.6\,\si{pc}$, and assuming the contribution of ejecta mass to $M_{\rm ISM}$ to be small, our estimate indicates a pre-explosion ISM density of around $ 1.1\times10^{-24}\,\si{g.cm^{-3}}$ when averaged over the present-day volume of Puppis A. Assuming an SNR age of $t\approx4000\,\si{yr}$, the expected explosion energy needed to produce a Sedov-Taylor shock wave \citep{Sedov59} equivalent to the observed SNR size is thus 
\begin{equation}
    E = \left(1.22 - 1.36 \right) \,\left ( \frac{r}{10.6\,\si{pc}} \right )^{9/2} 
    \left (    \frac{t}{4000 \,\si{yr}} \right )^{-2} \times 10^{51} \, \si{erg},   
\end{equation}
which is quite close to the canonical value of $10^{51} \, \si{erg}$.\footnote{The dependence $r^{9/2}$ arises from the combination of the mean density being proportional to $r^{-1/2}$ and the blast-wave energy being proportional to $r^{5}$.} The fact that this approach yields a realistic result is quite reassuring, concerning in particular the distance to Puppis A, as the previously accepted value of $2.2 \,\si{kpc}$ \citep{Reynoso03} would yield a factor $\sim 10$ larger energy. 
However, it is important to keep in mind that the assumptions made here are extremely crude, implying significant systematic uncertainties on the explosion energy. For instance, an explosion into an ISM of uniform density seems extremely unlikely, considering the strong brightness gradient between the northeast and southwest of Puppis A. Moreover, the shock wave of Puppis A is in reality clearly not perfectly spherical given its appearance on the sky. 

By combining our density estimate for each bin with the measured elemental abundances, it is straightforward to analogously estimate the masses of individual elements contributing to the observed emission in Puppis A. For those elements to which our fit is sufficiently sensitive, we obtain the following estimates: $M_{\rm O} \approx 0.29 \,M_{\odot}$, $M_{\rm Ne} \approx 0.085 \,M_{\odot}$, $M_{\rm Mg} \approx 0.032 \,M_{\odot}$, $M_{\rm Si} \approx 0.052 \,M_{\odot}$, $M_{\rm Fe} \approx 0.043 \,M_{\odot}$.  
While it is certain that a limited contribution to the emission of these elements comes from ejecta material, the major contribution to the observed mass most likely originates from the ISM. Therefore, these estimates probably represent the composition of the local ISM around Puppis A, whereas the composition of actual ejecta should be inferred directly from regions where these can be sufficiently isolated from their surroundings.

Therefore, to attempt to obtain mass estimates of individual ejecta clumps, we used our fit results given in Table \ref{DetailedFitsTable} for the ejecta-enhanced regions A$-$E  and proceeded analogously to our mass estimate for the whole remnant, with the modifications concerning assumed extent and electron-to-hydrogen ratio outlined in Sect.~\ref{DetailedFitSection}.
The resulting mass estimates, in total and for individual elements, are given in Table \ref{MassTable}. 

It can be immediately seen that approaches (i) and (ii) yield differences by up to an order of magnitude for both the total and element-specific masses. Therefore, making quantitative statements about the mass of individual ejecta knots is subject to large uncertainty, as both approaches are physically viable, and would likely require the spectral resolution of microcalorimeter detectors to be distinguishable \citep{Greco20}.
This is aggravated by additional large sources of systematic uncertainty such as the assumed three-dimensional extent of the respective feature, as well as the distance to Puppis A. 
This can be seen for regions C and D, where even after correcting for the different assumed distances,\footnote{The estimated mass depends on $d^{5/2}$, due to the contributions of emitting volume ($\propto d^{3}$) and density ($\propto d^{-1/2}$).} our mass estimates and those for analogous regions by \citet{Katsuda13} are discordant by a factor of a few, underlining the great importance of the assumed region volume for such computations. For instance, the oxygen mass $M_{\rm O}$ in the southern portion of the ejecta knot (region C) obtained in their study would correspond to $1.0\times10^{-4}\,M_{\odot}$ for normal, and $5.0\times10^{-4}\,M_{\odot}$ for pure-metal abundances, whereas we find values of $2.7\times10^{-4}\,M_{\odot}$ and $15\times10^{-4}\,M_{\odot}$, respectively. 

While not particularly constraining in an absolute sense, our analysis implies that most X-ray emitting ejecta mass in Puppis A is located in region E, which is the largest more or less contiguous region of strongly enhanced abundances in the SNR. It is somewhat unexpected that such a region would show a significant silicon enhancement, reaching an overall mass ratio $M_{\rm Si}/M_{\rm O} \sim 0.7$. 
Given the large size of this silicon-enriched region, and the compact nature of the more oxygen-rich ejecta knot (see Figs.~\ref{DiffImages} and \ref{TesselationImage}), it seems reasonable to conclude that the average composition of all X-ray emitting ejecta clumps is dominated by the elemental composition measured in region E. 
The implied mass ratio $M_{\rm Si}/M_{\rm O} \sim 0.7$ would be extremely high if it were to be seen as representative of the average composition of the nucleosynthetic products of the supernova, as numerical modelling tends to predict $M_{\rm Si}/M_{\rm O} \lesssim 0.2$ \citep[e.g.,][]{Sukhbold16,Tominaga07,Rauscher02} for the integrated yield of core-collapse supernova ejecta.
Since it is certain that Puppis A is the remnant of a core-collapse supernova, this apparent contradiction illustrates that,
even thousands of years after the explosion, individual ejecta elements, which originate from different layers of the progenitor star, have not fully mixed and exist in separate clumps. 
The apparent lack of light ejecta elements with respect to Si might be caused by an on average earlier reheating of the outer ejecta layers, rich in O, Ne, and Mg, by the reverse shock. 
This may mask the presence of a large fraction of light-element ejecta among macroscopic variations of elemental abundances across the SNR. Apart from the uncertain possibility of ejecta cooling out of the X-ray-emitting regime, the rapid destruction of compact ejecta clumps by the reverse shock (see Sect.~\ref{DiscEjDist}) may play a key role in this process, as it is expected to lead to stronger mixing with the ambient medium for ejecta material shocked earlier in time.
The heavy impact of these effects on the apparent composition of the observed ejecta material are the reason why any attempt to infer the progenitor mass of Puppis A based on the integrated composition of its X-ray emitting ejecta budget is likely flawed.

\section{Summary \label{Summary}}
At this time, the data set presented in this paper constitutes the deepest and highest-spatial-resolution X-ray observation of the entire Puppis A SNR taken with a single instrument. 
By dissecting the emission into narrow energy bands and performing spatially resolved spectral analysis, we have inferred the physical conditions and composition of the plasma across the entire SNR at hitherto unmatched sensitivity.

Our analysis has confirmed previous suggestions that parts of Puppis A, in particular in the southwest, are subject to strong foreground absorption \citep{Dubner13}. The overall variation of the absorption column over Puppis A was found to be at least of a factor five. Part of the absorption may be contributed on small spatial scales by dust within Puppis A, which is visible in the mid-infrared \citep{Arendt10}.  
Furthermore, we have shown that large-scale variations in plasma temperature span a range of around a factor two, ranging between around $0.35 \,\si{keV}$ in the western arc and $0.75\,\si{keV}$ at the northeast rim. Thanks to being comparatively close to CIE, the latter region exhibits the hardest X-ray emission in Puppis A, becoming dominant above $2\,\si{keV}$ in imaging. 
The combination of hot plasma in the northeast and enhanced absorption in the southwest is identified as the likely origin of the characteristic strip of hard emission crossing the SNR.

From the plasma ionization age and emission measure distributions, we have reconstructed the time of shock interaction of the emitting material, providing an interesting new look at Puppis A: several features, such as the northeast filament \citep{Katsuda10}, the ejecta knot \citep{Katsuda08,Hwang08} and the eastern edge of the BEK \citep{Hwang05} stand out prominently, as the material in all these regions appears to have recently interacted with a forward or reverse shock.   
No clear analogous signatures of recent shock interaction are observed in the western half of Puppis A, which may be related to the thinner ISM there. 

We have constructed elemental abundance maps of Puppis A, which reveal that elements typical for core-collapse SN ejecta (O, Ne, Mg, Si, S) only show a few concentrated enhancements in X-rays, which are spatially consistent with the location of ejecta enhancements identified by \citet{Katsuda08,Katsuda10} and \citet{Hwang08}. 
We have confirmed the spatially disjoint nature of light-element and IME ejecta, as the peak of oxygen, neon, and magnesium is clearly separated from the more spatially extended silicon and sulfur emission. None of the regions investigated are found to exhibit a convincing signature of enrichment with iron ejecta.  
The X-ray emission of the remainder of Puppis A generally implies solar or subsolar abundances for all elements, with only mild large-scale variations. The highest average abundances are found for neon, which may be 
related to the depletion of other elements onto dust grains \citep{Hwang08}, which would most strongly affect magnesium, silicon, and iron. 

An investigation of the distribution of silicon-rich ejecta with respect to the recoil direction of the neutron star yields a deviation of around $57^{\circ}$ from the expectation, significantly higher than previously measured \citep{Katsuda18}.
However, an important shortcoming of such a measurement is the unknown location of the reverse shock that may severely bias the observed distribution of ejecta material in Puppis A, which would also explain the apparent lack of ejecta clumps over the majority of the SNR. 

Our spatially resolved treatment of the emission has allowed us to calculate the mass of the swept-up ISM contributing to the observed X-ray emission of Puppis A, which we determine to around $78-87\,M_{\odot}$. We use this to crudely estimate the explosion energy needed to create a Sedov-Taylor blast wave equivalent to the approximate present-day radius of Puppis A, finding a value reasonably close to the canonical explosion energy of $10^{51} \, \si{erg}$. 
Furthermore, we have performed estimates of individual element masses in ejecta enhancements, finding that the largest ejecta-rich region exhibits an extremely high silicon-to-oxygen ratio, when compared to the expected integrated yield of a core-collapse supernova. 
This may be a further consequence of the biased view on the inherent ejecta composition in this mature SNR, as much of the light-element ejecta may have mixed with ISM or cooled below the X-ray emitting regime since their passage through the reverse shock.

After the completion of its four-year survey at the end of 2023, eROSITA will have mapped the entire X-ray sky at unprecedented sensitivity and spatial resolution. While the acquired exposure and statistics for SNRs in the all-sky survey will of course not be comparable to this study, we hope to have demonstrated the excellent capabilities of eROSITA regarding the detection and characterization of diffuse emission.
This will in particular be relevant for the search for new Galactic SNRs \citep[e.g.,][]{Churazov21,Becker21}, which may help reduce the discrepancy between the expected and observed number thereof. Furthermore, the survey will allow carrying out systematic studies of the spectral and morphological properties of the entire X-ray detected SNR population using a homogeneous data set (W. Becker et al., in prep.).

\begin{acknowledgements}
We are grateful to the anonymous referee for their detailed and helpful suggestions, which we believe have helped significantly improve the quality of our manuscript.  
We would like to thank N. Locatelli for fruitful discussions on X-ray absorption.  
Furthermore, we would like to thank M. Ramos-Ceja, I. Stewart, K. Dennerl, S. Friedrich, D. Coutinho, W. Kink, and I. Kreykenbohm for helpful input regarding instrument calibration and background, as well as data artifacts in the more recent data set of Puppis A.
MGFM acknowledges support by the International Max-Planck Research School on Astrophysics at the Ludwig-Maximilians University (IMPRS).  
\\
This work is based on data from eROSITA, the soft X-ray instrument aboard SRG, a joint Russian-German science mission supported by the Russian Space Agency (Roskosmos), in the interests of the Russian Academy of Sciences represented by its Space Research Institute (IKI), and the Deutsches Zentrum f\"ur Luft- und Raumfahrt (DLR). The SRG spacecraft was built by Lavochkin Association (NPOL) and its subcontractors, and is operated by NPOL with support from the Max Planck Institute for Extraterrestrial Physics (MPE). The development and construction of the eROSITA X-ray instrument was led by MPE, with contributions from the Dr. Karl Remeis Observatory Bamberg \& ECAP (FAU Erlangen-Nuernberg), the University of Hamburg Observatory, the Leibniz Institute for Astrophysics Potsdam (AIP), and the Institute for Astronomy and Astrophysics of the University of T\"ubingen, with the support of DLR and the Max Planck Society. The Argelander Institute for Astronomy of the University of Bonn and the Ludwig Maximilians Universit\"at Munich also participated in the science preparation for eROSITA.
The eROSITA data shown here were processed using the eSASS software system developed by the German eROSITA consortium.
\\
This research made use of Astropy,\footnote{\url{http://www.astropy.org}} a community-developed core Python package for Astronomy \citep{astropy:2013, astropy:2018}. Further, we acknowledge the use of the Python packages Matplotlib \citep{Hunter:2007}, SciPy \citep{SciPy}, and NumPy \citep{NumPy}.  
\end{acknowledgements}

\bibliographystyle{aa} 
\bibliography{Citations}

\onecolumn
\begin{appendix}

\section{Spectral parameter maps at lower-resolution binning \label{OtherMaps}}

\begin{figure*}[h!]
\centering
\includegraphics[width=18.0cm]{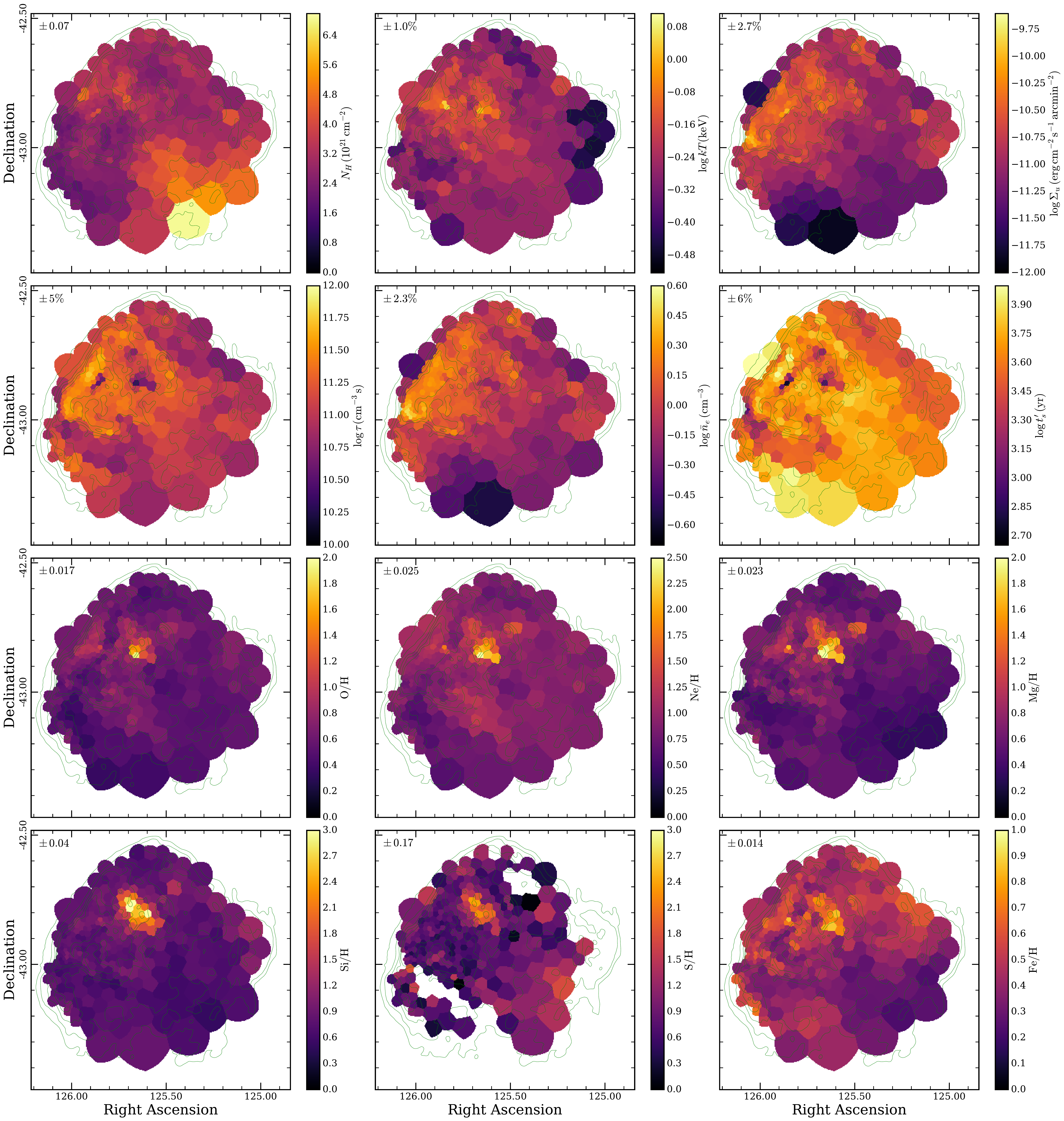} 
\caption{Same as Fig.~\ref{TesselationImage}, but with coarser binning due to a larger threshold of $S/N = 300$. In the map of $\rm S/H$, all bins with an absolute error larger than $0.4$ were masked.  
}
\label{TesselationImage_SNR300}
\end{figure*}

\begin{figure*}[h!]
\centering
\includegraphics[width=18.0cm]{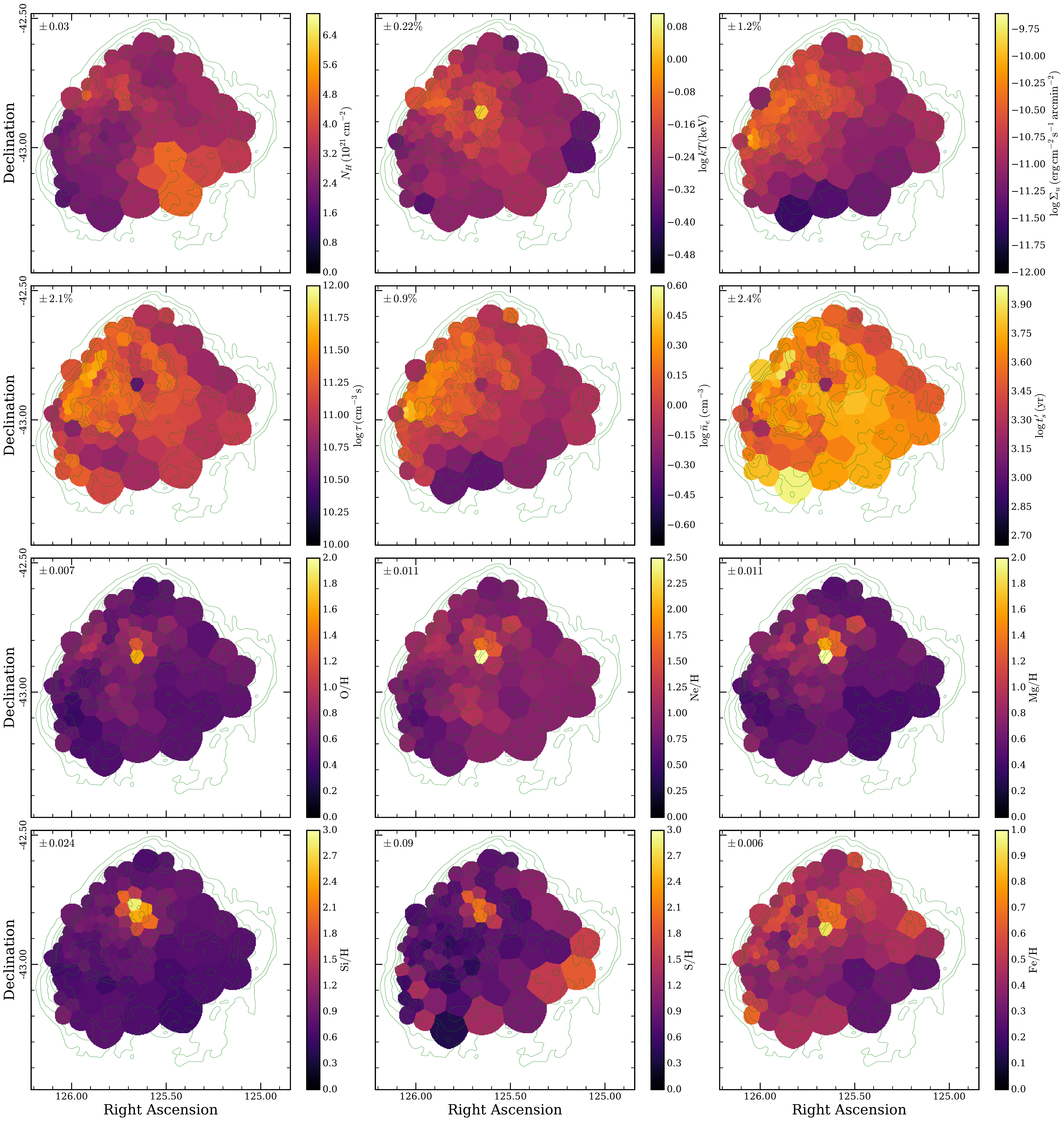} 
\caption{Same as Fig.~\ref{TesselationImage}, but with coarser binning due to a larger threshold of $S/N = 500$. 
}
\label{TesselationImage_SNR500}
\end{figure*}
\clearpage

\section{Distributions and correlations of spectral fit parameters \label{Correls}}

\begin{figure*}[h!]
\centering
\includegraphics[width=18.0cm]{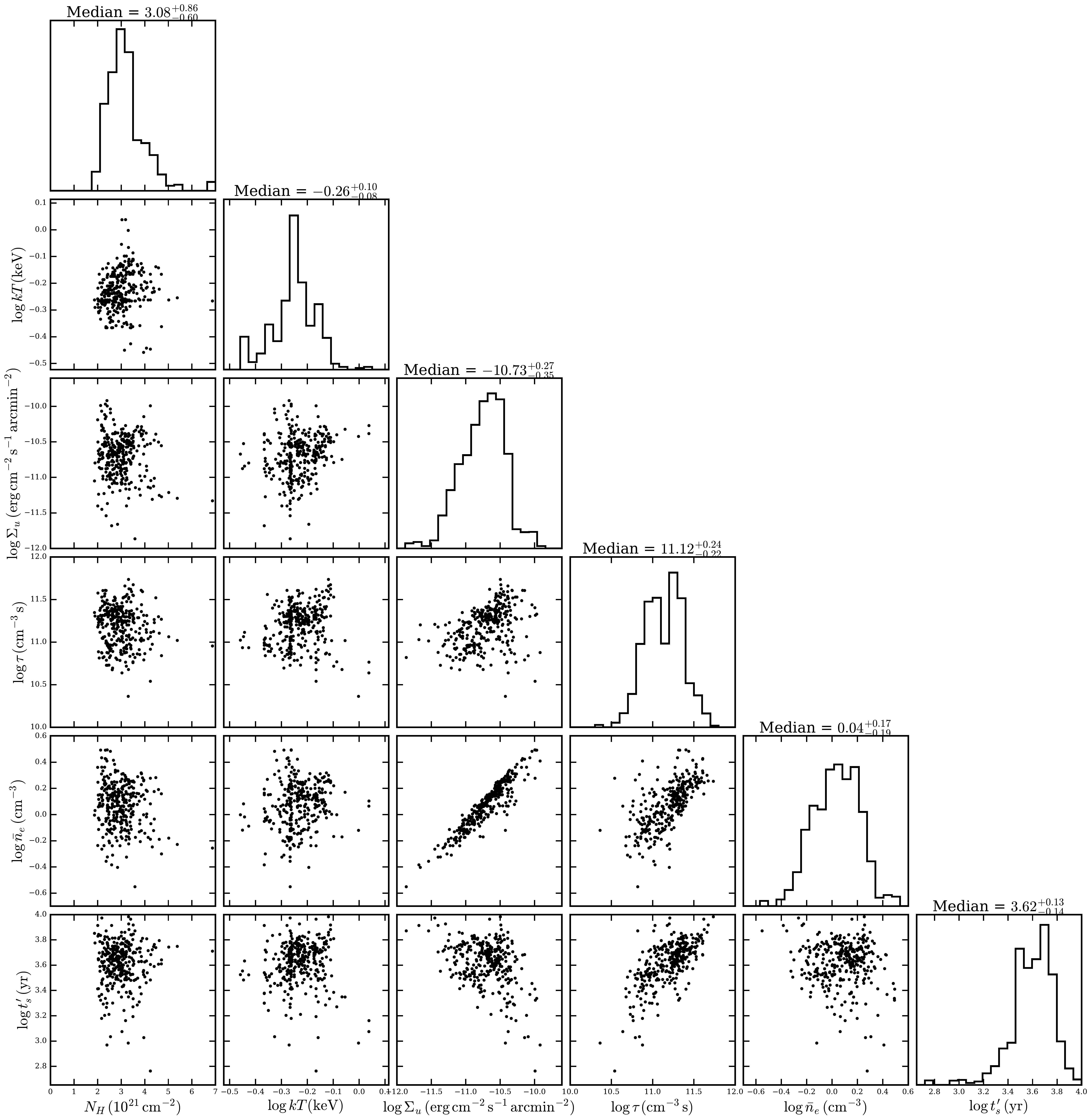} 
\caption{Correlation plot of the parameters describing foreground absorption and plasma conditions in the individual regions in Fig.~\ref{TesselationImage_SNR300}. 
We used the $S/N = 300$ binning here to reduce the statistical errors in the individual points, in order to suppress scatter in the resulting distributions. 
The distributions on the diagonal represent flux-weighted histograms of the respective parameter across the remnant, with the median and $68 \%$ central interval indicated above.}
\label{PlasmaCorrels}
\end{figure*}

\begin{figure*}[h!]
\centering
\includegraphics[width=18.0cm]{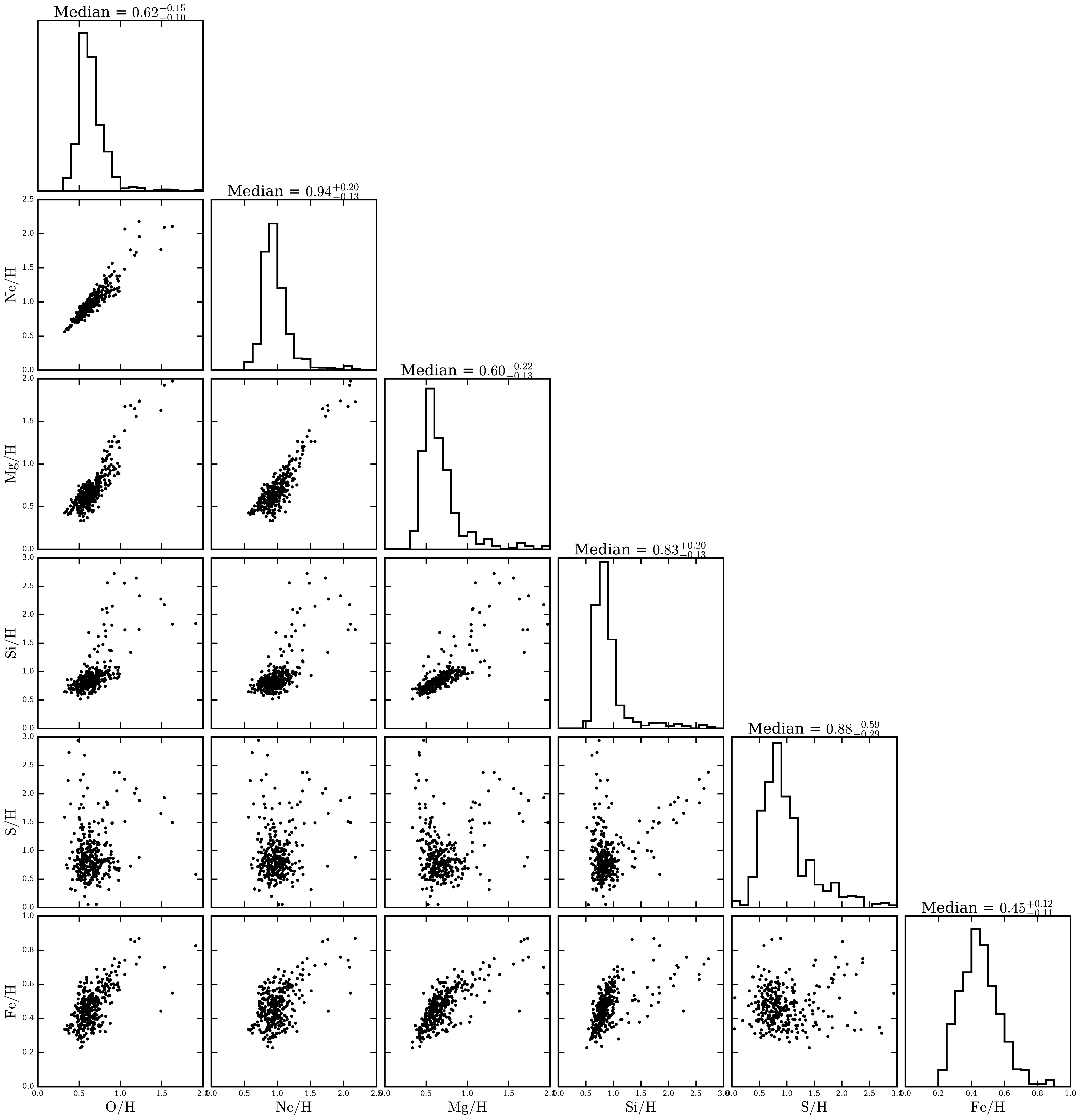} 
\caption{Same as Fig.~\ref{PlasmaCorrels}, but for the parameters describing elemental abundances. 
}
\label{AbundanceCorrels}
\end{figure*}

\begin{figure*}[h!]
\centering
\includegraphics[width=18.0cm]{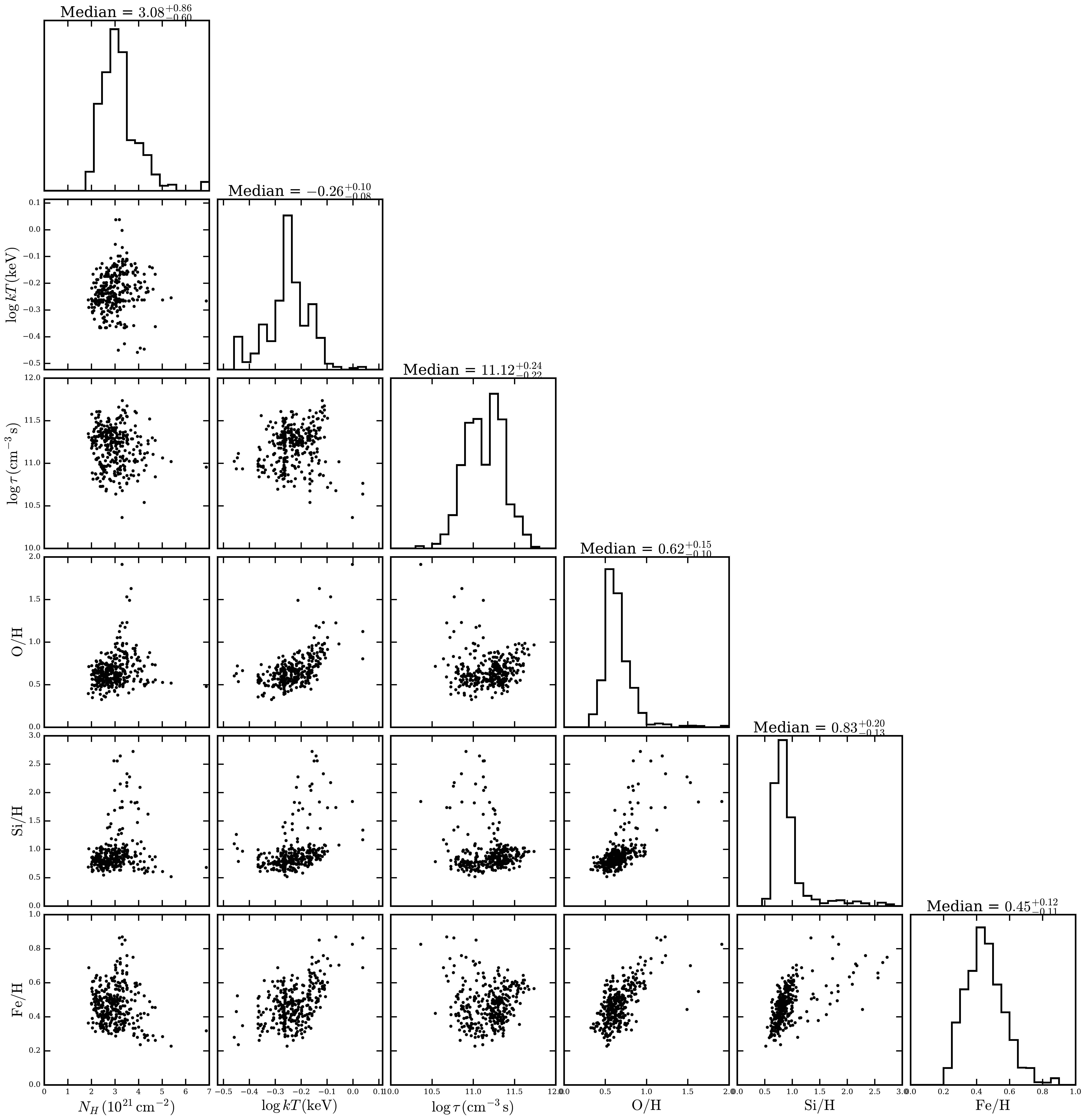} 
\caption{ Same as Fig.~\ref{PlasmaCorrels}, but illustrating the correlation between $N_{\rm H}$, $kT$, $\tau$ and the abundances of the characteristic elements O, Si, Fe. 
}
\label{MixedCorrels}
\end{figure*}

\end{appendix}

\end{document}